\documentclass[11pt,fleqn]{article}

\usepackage{amsfonts}
\usepackage{amsmath}
\usepackage{amssymb}
\usepackage{mathtools}
\usepackage{mathrsfs}
\usepackage{enumerate}
\usepackage{bbm}

\usepackage{pstricks}
\usepackage{subfig}
\usepackage{graphicx}
\usepackage{xcolor}
\usepackage{units}

\usepackage{caption}
\usepackage{booktabs}
\usepackage{textcomp}
\usepackage{array}
\usepackage{cite}

\date{\today}

\newcommand{\del}{\partial}                            
\newcommand{\ol}[1]{\overline{#1}}                     
\newcommand{\abs}[1]{ \left| #1 \right| }              
\newcommand{\klammer}[1]{\left(#1\right)}              
\newcommand{\fkt}[1]{\operatorname{#1}\,}              
\newcommand{\real}{\Re\mathfrak{e}\hspace{1pt}}        
\newcommand{\s}[1]{\texttt{#1}}                        
\newcommand{\cbox}[1]{{\hspace{1pt}\fboxsep0pt\fbox{\colorbox{#1}{\phantom{$\overline{aaa}$}}}}}

\definecolor{brownochre}{HTML}{94471E}

\newcolumntype{z}[1]{>{\RaggedRight\hspace{0pt}}p{#1}}
\newcolumntype{w}[1]{>{\RaggedRight\hspace{0pt}}p{#1}}
\newcolumntype{v}[1]{>{\Centering\hspace{0pt}}p{#1}}

\newcommand{\chargino}{\widetilde{\chi}^+}
\newcommand{\neutralino}{\widetilde{\chi}^0}

\DeclareCaptionLabelSeparator{mysep}{\hspace{3pt}:\hspace{3pt}}
\DeclareCaptionLabelFormat{mypiccap}{Fig.\hspace{3pt}{#2}}
\DeclareCaptionLabelFormat{mytabcap}{Tab.\hspace{3pt}{#2}}

\begin{document}

\date{}
\title{
\vskip 2cm
{\bf\huge Mirage Pattern from the Heterotic String}\\[0.8cm]}

\author{{\sc\normalsize
Val\'eri L\"owen\footnote{E-mail: loewen@th.physik.uni-bonn.de} and Hans Peter Nilles\footnote{E-mail: nilles@th.physik.uni-bonn.de}\!
\!}\\[1cm]
{\normalsize Physikalisches Institut der Universit\"at Bonn}\\
{\normalsize Nussallee 12, 53115 Bonn, Germany}\\[1cm]
}
\maketitle \thispagestyle{empty}
\begin{abstract}
{We provide a simple example of dilaton stabilization
in the framework of heterotic string theory. It requires
a gaugino condensate and an up-lifting sector similar to
the one postulated in type IIB string theory. 
Its signature is a hybrid mediation of
supersymmetry breakdown with a variant of a mirage pattern 
for the soft breaking terms. The set-up is suited for the
discussion of heterotic MSSM candidates.
}
\end{abstract}

\clearpage

\tableofcontents
\newpage

\section{Introduction}
\label{sec:introduction}

With the recent success of model building in the framework of the heterotic
$E_8 \times E_8$ string theory 
\cite{Lebedev:2006kn,Lebedev:2007hv,Choi:2006qh} 
it becomes important to reconsider the
questions of moduli stabilization and supersymmetry breakdown.
Early attempts considered fluxes and gaugino condensates 
\cite{Derendinger:1985kk,Dine:1985rz,Derendinger:1985cv},
race track superpotentials 
\cite{Krasnikov:1987jj,Casas:1990qi} 
and/or K\"ahler stabilization 
\cite{Casas:1996zi,Binetruy:1996xja,Barreiro:1997rp}.

More recently, these questions have been studied in the framework
of type IIB string theory. While explicit model building towards the
MSSM (the minimal supersymmetric extension of the standard model)
is more difficult in this framework, it is usually argued that
the stabilization of moduli can be achieved quite easily. The reason
is the appearance of two types of fluxes [Neveu-Schwarz--Neveu-Schwarz (NS-NS) and Ramond--Ramond (R-R)] \cite{Dasgupta:1999ss,Giddings:2001yu}
while only one of them is present in the heterotic theory.\footnote{More fluxes can appear if we go beyond Calabi-Yau
compactification\cite{Becker:2003yv,LopesCardoso:2003sp,Gurrieri:2004dt}.} With the inclusion of nonperturbative effects like gaugino condensation one might 
then stabilize all moduli \cite{Kachru:2003aw}. Still, there is a problem with the 
adjustment of the vacuum energy $E_\s{VAC}$ to a small positive value. One has to
postulate a so-called ``up-lifting'' sector that adjusts $E_\s{VAC}$ to the
desired value \cite{Kachru:2003aw}. It turns out that this up-lifting sector has important
consequences for the explicit pattern of supersymmetry breakdown 
\cite{Choi:2004sx}. Instead of modulus mediation one is led to a hybrid mediation scheme and
a so-called ``mirage pattern'' of the soft breaking terms emerges 
\cite{Choi:2005ge,Choi:2005uz,Endo:2005uy,Falkowski:2005ck,Baer:2006id}.

Within the framework of the heterotic string, the importance of this
up-lifting sector has not been fully appreciated so far. The present paper is an 
attempt to fill this gap. To illustrate the importance of this sector, we consider a simple
example: a gaugino condensate in the absence of a flux background. This is known to
lead to a so-called run-away potential with a supersymmetric vacuum at 
$S\rightarrow +\infty$ (where $S$ denotes the dilaton field). Attempts to stabilize the dilaton 
at a finite value include the consideration of nonperturbative correction to the K\"ahler
potential \cite{Casas:1996zi,Binetruy:1996xja,Barreiro:1997rp}. 
The resulting scheme is believed to provide a mediation 
of supersymmetry breakdown that is dominated by the dilaton. Still the 
question of the fine-tuning of the vacuum energy has to be addressed. In the existing examples it
seems again that an additional sector is needed to adjust the vacuum energy
to the desired value \cite{Casas:1996zi}, just like the up-lifting sector in the type IIB
case. To be explicit, we consider the model of Barreiro, Carlos and Copeland (BCC) ref. \cite{Barreiro:1997rp}
that appears to be particularly suited for a class of realistic heterotic models 
\cite{Lebedev:2006tr}. A stabilization of the dilaton can be
achieved, but the actual value of $E_\s{VAC}$ turns out to be large and 
positive so that the desired ``up-lifting'' mechanism should provide a ``down-lift'' of 
$E_\s{VAC}$. We here consider the mechanism of Lebedev, Nilles and Ratz (LNR) \cite{Lebedev:2006qq} to adjust $E_\s{VAC}$. In the 
framework of type IIB theory this is known as $F$-term uplifting 
\cite{GomezReino:2006dk,Lust:2006zg,Dudas:2006gr,Abe:2006xp,Lebedev:2006qc,Serone:2007sv} 
and could be originated in dynamical schemes as considered in \cite{Intriligator:2006dd}
or schemes that might require the existence of additional branes and/or anti-branes \cite{Kachru:2003aw}. 
An explicit discussion of supersymmetry breakdown leads to similar conclusions as in the type IIB case, the
uplifting sector is important for the mechanism of supersymmetry (SUSY) breakdown and its mediation such 
that a variant of a mirage pattern emerges \cite{Lebedev:2006qq}. There are, however, some quantitative 
differences between the heterotic and the type IIB case that will be discussed in detail.

So far it seems that the stabilization of the dilaton requires
sizeable nonperturbative corrections to the K\"ahler potential 
while the uplifting sector adjusts $E_\s{VAC}$. A closer inspection,
however, reveals the surprising fact that one can switch off the
corrections to the K\"ahler potential and still remain with a
stabilized dilaton! 
{\emph{The uplifting sector alone is responsible for
both, modulus stabilization and adjustment of the vacuum energy.}}
Stabilization of the dilaton neither requires nonvanishing flux
nor corrections to the K\"ahler potential. The appearance of
a gaugino condensate, however, remains crucial.

This remarkable fact shows that modulus stabilization in heterotic
string theory is not as difficult as usually assumed. One just
needs a suitable uplifting sector very similar to the one postulated
in the framework of type IIB string theory. It also shows the
importance of this uplifting sector for moduli stabilization
and supersymmetry breakdown. Moreover it leads to a hybrid mediation
scheme and its signature is 
a mirage pattern of the soft supersymmetry breaking terms.

The paper is organized as follows. In section 2 we shall discuss the
model of 
ref \cite{Barreiro:1997rp} with a gaugino condensate and nonperturbative
corrections to the K\"ahler potential. We present the
heterotic analogy of uplifting in the spirit of 
LNR \cite{Lebedev:2006qq}, the 
adjustment of the vacuum energy of the metastable vacuum and the
discussion of supersymmetry breakdown. This will be followed by
a discussion of the soft supersymmetry breaking parameters,
the appearance of a mirage pattern and the phenomenological
properties of the set-up. Section 4 will contain some concluding
remarks.

\section{Dilaton stabilization in heterotic string theory}
\label{sec:theory}

Dilaton stabilization in the context of heterotic string theory can 
occur via a racetrack mechanism \cite{Krasnikov:1987jj,Casas:1990qi} 
or by the means of nonperturbative corrections to the K\"ahler potential 
\cite{Casas:1996zi,Binetruy:1996xja,Barreiro:1997rp}. The former option, 
consisting of at least  two gaugino condensates, 
leads to a scenario where both the auxiliary field of the 
dilaton and the $T$ modulus are nonzero with typically $F^T>F^S$. 
The latter option might require just one gaugino condensate, but 
sizeable nonperturbative corrections to the tree level K\"ahler potential. 
Under certain circumstances this leads to a scenario with $F^T=0$ which 
is also known as the \emph{dilaton domination scenario}.

Casas \cite{Casas:1996zi} has investigated the impact of an arbitrary 
K\"ahler potential on the low energy theory and also introduced 
possible string motivated nonperturbative corrections to the 
K\"ahler potential. The distinct feature of this ansatz was the fact 
that it became possible to stabilize the dilaton at phenomenologically 
acceptable values $\real{S}\simeq 2$, although the minimum of the scalar 
potential turned out to be de Sitter.

In this work we shall investigate the impact of the presence of an
up-lifting sector (consisting of hidden sector matter fields
\cite{Lebedev:2006qq})
on a set-up with just one gaugino condensate and the 
tree-level K\"ahler potential. But let us first repeat the arguments
of \cite{Casas:1996zi}
and present the explicit model of BCC \cite{Barreiro:1997rp}.

\subsection{Problems with a single gaugino condensate}
\label{sub:difficulty}

Consider \cite{Casas:1996zi,Barreiro:1997rp}
\begin{align}
W(S,T) &= -A\frac{1}{\eta^6(iT)}e^{-a\,S},
\label{eqn:super}
\end{align}
where $\eta(iT)$ is the Dedekind eta function (convention as in \cite{Polchinski:1998rq}), insuring the correct transformation under the $SL(2,\mathbbm{Z})$ target space modular invariance and $a=\nicefrac{8\pi^2}{N}$ if the condensing gauge group is $SU(N)$. The K\"ahler potential at tree-level is given by
\begin{align}
K = -3\log\klammer{T+\ol{T}} - \log\klammer{S+\ol{S}}.
\label{eqn:kahler}
\end{align}
In the supergravity language the scalar potential and the $F$ terms are expressed in terms of the function
\begin{align}
G &= K + \log W\ol{W},
\label{eqn:gfunc}
\end{align}
with $K$ and $W$ being the K\"ahler potential and the superpotential, respectively. The scalar potential is given by
\begin{align}
V &= e^{G}\, \klammer{G^{-1}_{\alpha\ol{\beta}}\, G_\alpha\, G_{\ol{\beta}} - 3},
\label{eqn:scalarpot}
\end{align}
where $\alpha$, $\beta$ denote differentiation with respect to the fields and $G^{-1}_{\alpha\ol{\beta}}=K^{-1}_{\alpha\ol{\beta}}$ is the inverse K\"ahler metric. The $F$ terms are found to be
\begin{align}
F^\alpha &= e^{G/2}\, G^{-1}_{\alpha\ol{\beta}}\, G_{\ol{\beta}}.
\label{eqn:fterms}
\end{align}
Using eqs.\,(\ref{eqn:super},\,\ref{eqn:kahler}) we obtain
\begin{align}
F^S &= e^{G/2}\, K^{-1}_{S\ol{S}}\, \klammer{K_{\ol{S}} - a }, \label{eqn:oldfs}\\
F^T &= - e^{G/2}\, \klammer{T+\ol{T}}^2 \mathcal{E}(T,\ol{T})\label{eqn:oldft}
\end{align}
with $\mathcal{E}(T,\ol{T})=(T+\ol{T})^{-1}+2\eta^{-1}(i T)\del\eta(i T)/\del T$. In this work we will assume that the K\"ahler modulus is stabilized 
at one of the self-dual points of the $\mathcal{E}$ function, leading 
to $F^T=0$. From now on we will drop the $T$ dependence in 
K\"ahler potential, gauge kinetic function and superpotential 
and rescale $A\,\eta^{-6}(i T_0)\rightarrow A$.

In order to provide the formation of a minimum and so to stabilize the dilaton we look at the stationary point condition
\begin{align}
V_S &= G_S\, V + e^G \, \frac{\del}{\del S}\klammer{K^{-1}_{S\ol{S}}\, G_{S}\, G_{\ol{S}}} \stackrel{!}{=}0,
\label{eqn:oldstationary}
\end{align}
which relates the derivatives of $G$ as
\begin{align}
K^{-1}_{S\ol{S}}\,G_S\,\abs{G_S}^2 - K_{S\ol{S}S}\, K^{-2}_{S\ol{S}}\,\abs{G_S}^2-G_S &= 0.
\label{eqn:oldgrelation}
\end{align}
If we only use the tree-level K\"ahler potential $-\log\klammer{S+\ol{S}}$, eq.\,(\ref{eqn:oldgrelation}) becomes
\begin{align}
\left[ \klammer{2\real{S}}^2 \abs{\frac{1}{2\real{S}}+a}^3 - 4{\real{S}}\abs{\frac{1}{2\real{S}}+a}^2 - \klammer{\frac{1}{2\real{S}}+a}\right]_{\real{S}\simeq 2} &= 0,
\label{eqn:oldwrong}
\end{align}
which can not be solved for any reasonable value of $a$ or $N$, respectively. 
\begin{figure}
\captionsetup[figure]{labelfont={footnotesize,bf},textfont=footnotesize,labelsep=mysep,labelformat=mypiccap,format=default,justification=RaggedRight,width=7cm,indent=5pt}
\begin{minipage}{0.5\linewidth}
\includegraphics[width=\linewidth]{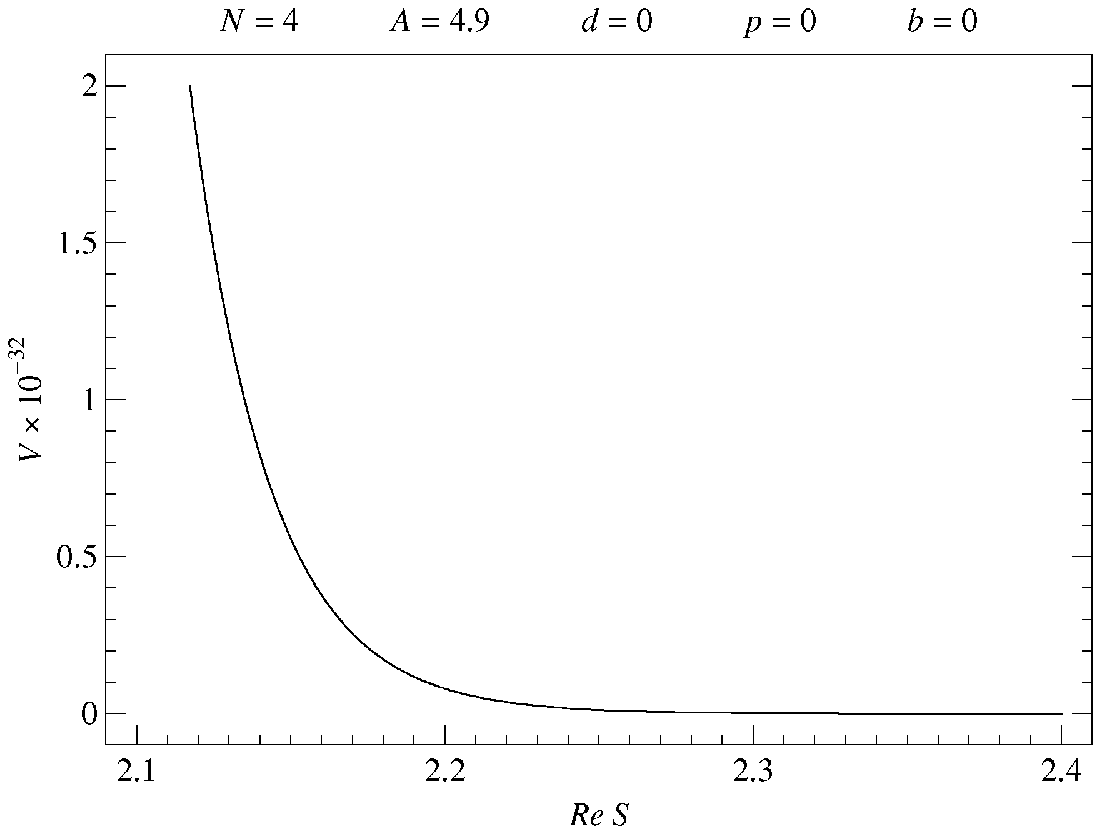}
\caption{Dilaton scalar potential for one condensate and the tree-level K\"ahler potential.}
\label{fig:runaway}
\end{minipage}
\begin{minipage}{0.5\linewidth}
\includegraphics[width=\linewidth]{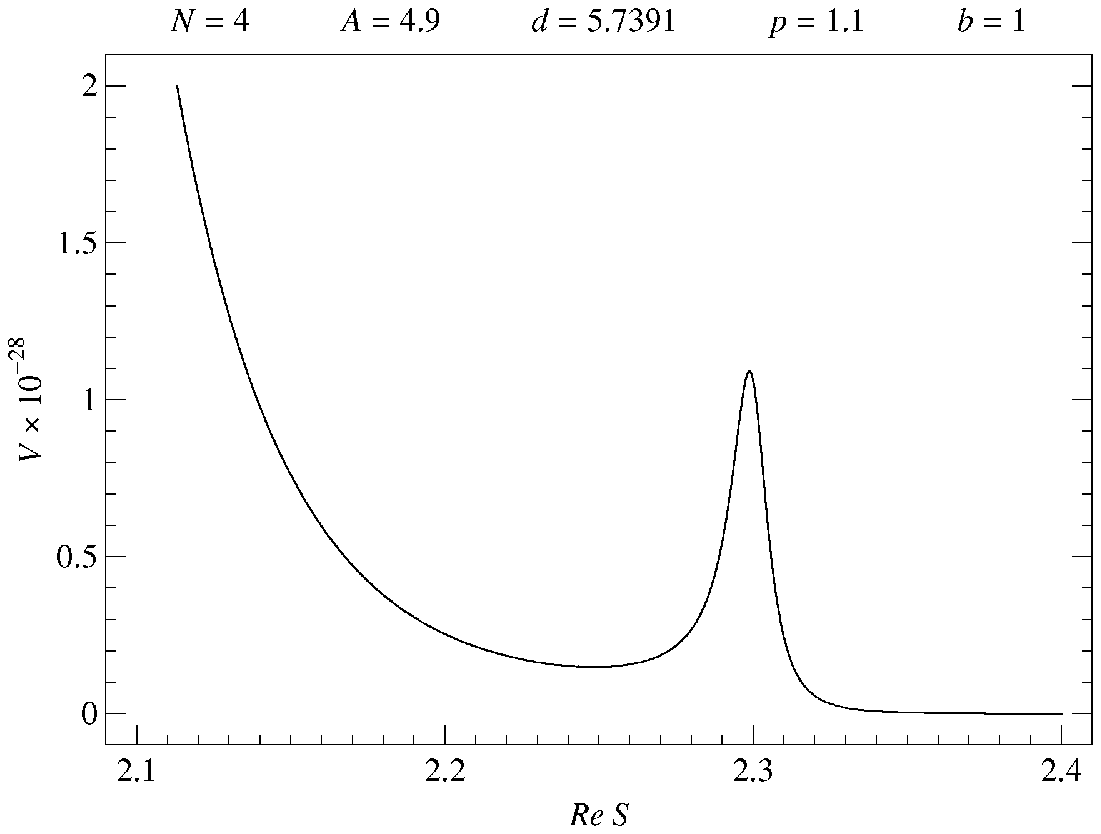}
\caption{Dilaton scalar potential for one condensate and K\"ahler potential as given by eq.\,(\ref{eqn:npkahler}).}
\label{fig:copeland}
\end{minipage}
\end{figure}
Therefore one obtains a runaway potential for the dilaton (fig.\,\ref{fig:runaway}). Adding nonperturbative corrections to standard tree-level K\"ahler potential as motivated in \cite{Casas:1996zi}
\begin{align}
K_{\texttt{TREE}+\texttt{NP}} = \log\left[ \frac{1}{S+\ol{S}} + d\, \left(\frac{S+\ol{S}}{2}\right)^{p/2} e^{-b\sqrt{\frac{S+\ol{S}}{2}}} \right]
\label{eqn:npkahler}
\end{align}
provides an extra contribution to eq.\,(\ref{eqn:oldwrong}) which also must be sizeable in order to allow for a solution. The explicit model of \cite{Barreiro:1997rp} typically obtains a potential
as displayed in fig.\,\ref{fig:copeland}.

Stabilization of the dilaton depends
crucially on the presence of nonperturbative corrections to the 
K\"ahler potential. Otherwise we would have a run-away behaviour of
the dilaton and no stabilization.
Typically the vacuum energy turns
out to be too large and a ``down-lift'' to a smaller value is necessary. 
An additional contribution to the scalar potential is 
therefore needed to 
give a solution to eq.\,(\ref{eqn:oldgrelation}). The role of such an
additional contribution should allow the adjustment of the vacuum energy
to a zero (or small) value. \emph{We shall see that such an additional sector
will stabilize the dilaton even in the absence of the nonperturbative
corrections to the K\"ahler potential.}

\subsection{Down-lifting the vacuum energy}
\label{sub:combination}

We would like to investigate the interaction of the set-up described 
in the previous section with a new hidden sector designed to 
adjust the vacuum energy. 
This is the heterotic analogy of \cite{Lebedev:2006qq}. 
The corresponding K\"ahler potential reads
\begin{align}
K = -\log\klammer{ {S+\ol{S}} } + C\ol{C},
\label{eqn:newkahler}
\end{align}
with $C$ being a hidden sector matter field, assumed to be a singlet under unbroken gauge symmetries. From eq.\.(\ref{eqn:scalarpot}) we obtain
\begin{align}
V &= e^{G}\, \klammer{K^{-1}_{S\ol{S}}\, G_S\, G_{\ol{S}} + G_C\,G_{\ol{C}} - 3}.
\label{eqn:newpot}
\end{align}
As obvious from eq.\,(\ref{eqn:newpot}) there are now new contributions to the scalar potential which, by properly adjusting the parameters of the new sector, may modify the shape of the scalar potential.

Let us for concreteness and simplicity
consider the Polonyi superpotential \cite{Polonyi:1977pj} and focus on the possibility of having a Minkowski vacuum. The superpotential for this choice is given by
\begin{align}
W &= \omega + \mu^2\, C -A\,e^{-a\, S},
\label{eqn:newsuper}
\end{align}
where $\omega$ and $\mu^2$ are constants. The scalar potential eq.\,(\ref{eqn:newpot}) becomes
\begin{align}
V &= e^{K}\, \klammer{ \abs{W_S \klammer{S+\ol{S}}-W}^2 + \abs{W\,\ol{C}+W_C}^2 - 3\abs{W}^2} \\
  &\equiv e^K \klammer{ \abs{\gamma^S}^2 + \abs{\gamma^C}^2 - 3\abs{W}^2 },
\label{eqn:newpot2}
\end{align}
where we have introduced the quantities
\begin{align}
\gamma^S &= \ol{W}_{\ol{S}} \klammer{S+\ol{S}}-\ol{W} \label{eqn:gammas},\\
\gamma^C &= W\,\ol{C}+W_C \label{eqn:gammac},
\end{align}
which denote the contribution of the dilaton and hidden sector matter 
field to SUSY breaking, respectively. 
The stationary point conditions now become
\begin{align}
\frac{\del V}{\del S} &= -\frac{V}{S+\ol{S}} + e^K \Big[ \ol{W}_{\ol{S}}\,\ol{\gamma}^S + W_{SS}\klammer{S+\ol{S}} + W_S\,\ol{C}\,\ol{\gamma}^C - 3W_S\ol{W} \Big]\stackrel{!}{=}0, \label{eqn:stationarys}\\
\frac{\del V}{\del C} &= V\,\ol{C} + e^K \Big[ -W_C\,\gamma^S + \ol{W}\,\ol{\gamma}^C + W_C\,\ol{C}\,\gamma^C -3W_C\ol{W} \Big]\stackrel{!}{=}0. \label{eqn:stationaryc}
\end{align}
In the following we show that eqs.\,(\ref{eqn:stationarys},\,\ref{eqn:stationaryc}) are satisfied at
\begin{align}
S = S_0 = \mathcal{O}(1), \quad C = C_0 = \mathcal{O}(1).
\label{eqn:vev}
\end{align}
To prove whether the stationary point corresponds to a minimum we have to calculate the Hessian matrix for the scalar potential and subsequently evaluate the eigenvalues. In this analysis we can neglect the vacuum energy $V_0=V(S_0,C_0)\lll 1$. Up to the factor $\exp{\left[K(S_0,C_0)\right]}\sim\mathcal{O}(1)$ the second derivatives of $V$ are given by
\begin{align}
V_{SS} &\simeq W_{SSS}\klammer{S_0+\ol{S}_0}\gamma^S + 2W_{SS}\klammer{S_0+\ol{S}_0}\ol{W}_{\ol{S}} + W_{SS}\klammer{\gamma^S + \ol{C}_0\gamma^C - 3\ol{W}_0}, \label{eqn:vss}\\
V_{S\ol{S}} &\simeq \abs{W_{SS}\klammer{S_0+\ol{S}_0}}^2 + W_{SS}\gamma^S + \ol{W}_{\ol{S}\ol{S}}\ol{\gamma}^S + \abs{W_S}^2\klammer{-2+\abs{C_0}^2}, \label{eqn:vssbar}\\
V_{CC} &\simeq 2\ol{W}_0\ol{C}_0 W_C, \label{eqn:vcc}\\
V_{C\ol{C}} &\simeq \abs{W_0 + W_C C_0}^2, \label{eqn:vccbar}\\
V_{SC} &\simeq -\ol{W}_{\ol{S}}W_C + \ol{W}_0 W_S \ol{C}_0, \label{eqn:vsc}\\
V_{S\ol{C}} &\simeq -\ol{W}_{\ol{C}} W_{SS}\klammer{S_0+\ol{S}_0} + W_S\klammer{\ol{W}_{\ol{C}}\abs{C_0}^2 + \gamma^C -3\ol{W}_{\ol{C}}}. \label{eqn:vscbar}
\end{align}
One can now consider two cases. For $\gamma^C\ll\gamma^S$ the SUSY breaking is 
dominated by the dilaton. 
However this does not correspond to a satisfactory choice as the contribution 
from the Polonyi sector would then not be sufficient to solve eqs.\,(\ref{eqn:stationarys},\,\ref{eqn:stationaryc}). Therefore it is more interesting to look at the so-called \emph{matter dominated} SUSY breaking where $\gamma^S\ll\gamma^C$. The leading order term in eqs.\,(\ref{eqn:vss}--\ref{eqn:vscbar}) is $\abs{W_{SS}(S_0+\ol{S}_0)}^2$. Next to leading order terms are $\abs{W_{SS}(S_0+\ol{S}_0)\ol{W}_{\ol{C}}}$ and $\abs{W_0}^2$. All other contributions are sub-leading and can be neglected in this analysis.\footnote{The term $W_{SS}(S_0+\ol{S}_0)\ol{W}_{\ol{S}}$ in the matrix element $V_{SS}$ contributes at the next to leading order. We however neglect it here to keep the Hessian as simple as possible in order to obtain transparent equations for the eigenvalues.}

\noindent{}Thus we arrive at
\begin{align}
\frac{\del^2 V}{\del x_i\del x_j} &\sim \left(%
\begin{array}{cccc}
       \abs{\Gamma}^2 &                   0 & \Gamma\,\ol{\theta} & 0 \\
                    0 &      \abs{\Gamma}^2 &                   0 & \ol{\Gamma}\,\theta \\
  \ol{\Gamma}\,\theta &                   0 &              \Delta & 0 \\
                    0 & \Gamma\,\ol{\theta} &                   0 & \Delta \\
 \end{array}%
\right),
\label{eqn:hessian}
\end{align}
with
\begin{align}
\Gamma &= W_{SS}\klammer{S_0+\ol{S}_0}, \\
\theta &= -W_C,\\
\Delta &= \abs{W_0}^2,
\end{align}
where $\abs{\Gamma}\gg\abs{\theta},\Delta$. The indices of the Hessian are defined as $(x_1,x_2,x_3,x_4)=(S,\ol{S},C,\ol{C})$. The eigenvalues are given by
\begin{align}
\frac{1}{2}\klammer{ \Delta + \abs{\Gamma}^2 - \sqrt{\Delta^2-2\Delta\abs{\Gamma}^2 +\abs{\Gamma}^4 + 4\abs{\Gamma}^2\abs{\theta}^2 } } &\simeq \frac{\Delta}{2}, \label{eqn:ev1} \\ 
\frac{1}{2}\klammer{ \Delta + \abs{\Gamma}^2 + \sqrt{\Delta^2-2\Delta\abs{\Gamma}^2 +\abs{\Gamma}^4 + 4\abs{\Gamma}^2\abs{\theta}^2 } } &\simeq \frac{\Delta}{2} + \abs{\Gamma}^2
\label{eqn:ev2}
\end{align}
and are all positive. This proves that the stationary point is a local minimum. Moreover the spectrum consists of two heavy states with masses of order $\abs{\Gamma}\sim\abs{W_{SS}}$ and two light states with masses of order $\sqrt{\Delta}=\abs{W_0}\sim\mu^2$.

\subsection{Adjusting the cosmological constant}
\label{sub:minkowski}

\begin{figure}
\captionsetup[figure]{labelfont={footnotesize,bf},textfont=footnotesize,labelsep=mysep,labelformat=mypiccap,format=default,justification=RaggedRight,width=7cm,indent=5pt}
\begin{minipage}{0.5\linewidth}
\includegraphics[width=\linewidth]{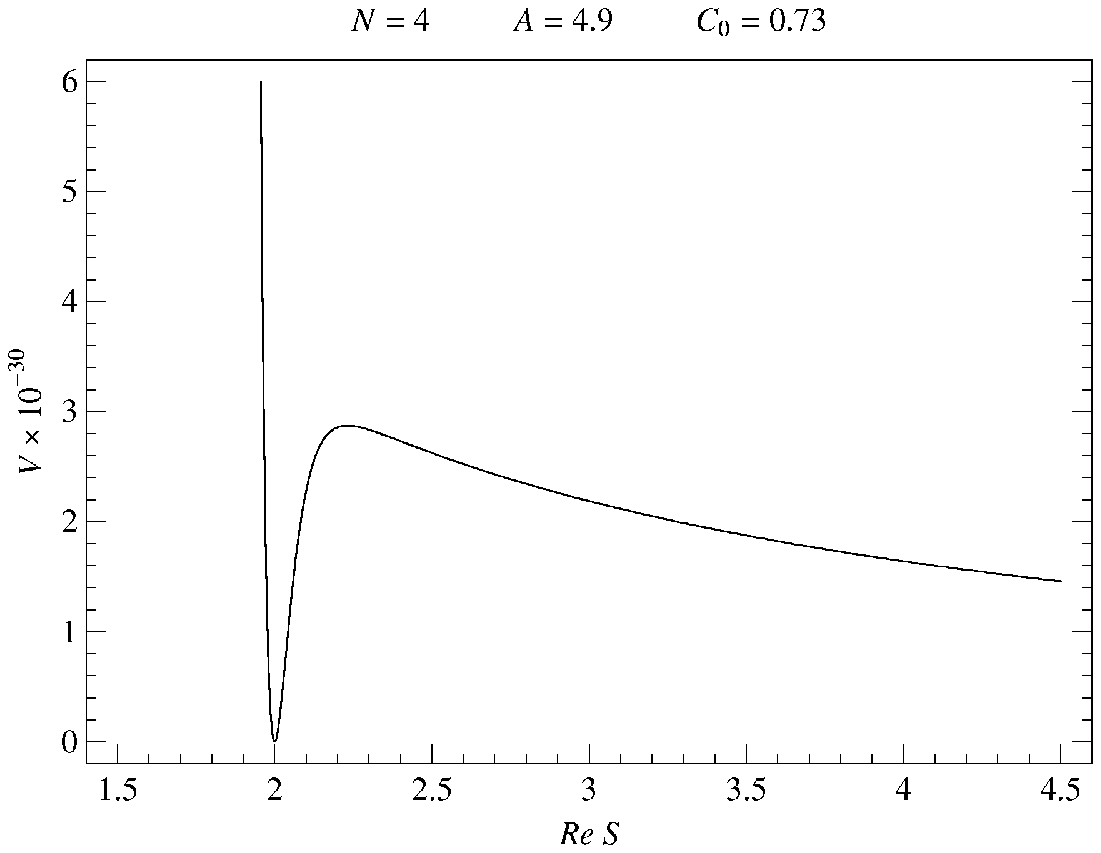}
\caption{Scalar potential for one condensate and a Polonyi field.}
\label{fig:sdirection}
\end{minipage}
\begin{minipage}{0.5\linewidth}
\includegraphics[width=\linewidth]{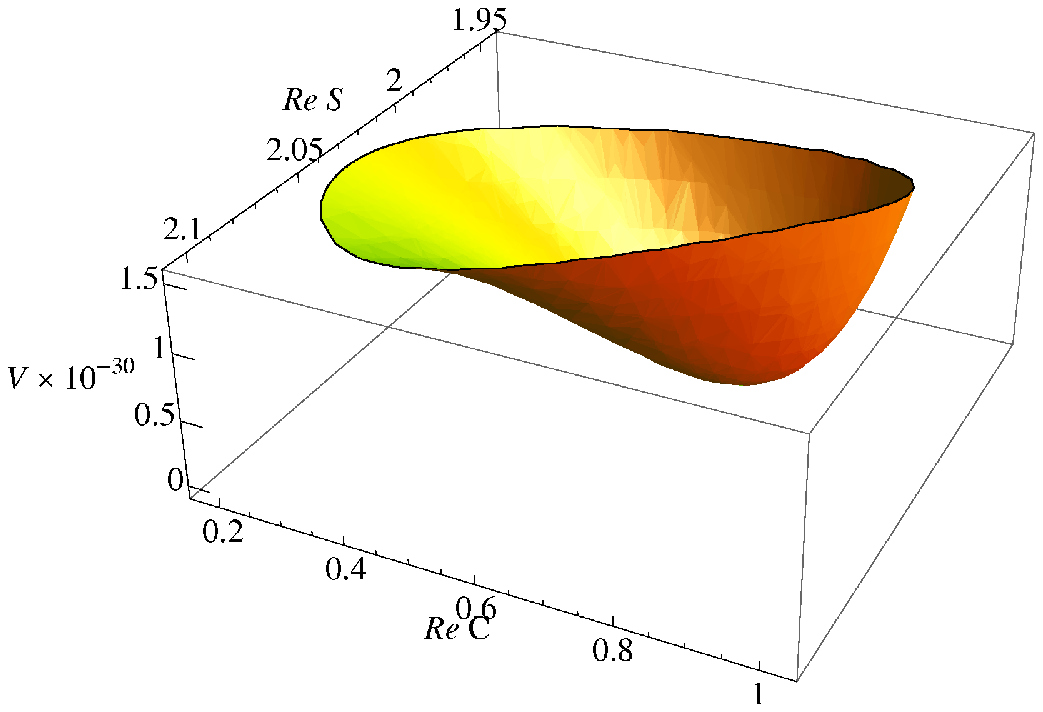}
\caption{Total scalar potential in $S$ and $C$ direction with Minkowski minimum at $S_0=2$.}
\label{fig:3dplot}
\end{minipage}
\end{figure}

In consideration of a vanishing/small cosmological constant eq.\,(\ref{eqn:newpot}) can be parameterized as 
\begin{align}
G_C\,G_{\ol{C}} + \lambda &= 3+\epsilon,
\label{eqn:estmation1}
\end{align}
where $\epsilon$ is a fine-tuning parameter and $\lambda=K^{-1}_{S\ol{S}}\, G_S\, G_{\ol{S}}\ll 1$. Then the vacuum energy is
\begin{align}
V(S_0) &= V_0 \sim \epsilon\mu^4.
\label{eqn:estimation2}
\end{align}
The solution to first order in $\epsilon$ reads
\begin{align}
\omega &\simeq \klammer{2 - \sqrt{3-\lambda} - \frac{\epsilon}{\sqrt{3-\lambda}}}\mu^2, \label{eqn:estimation3}\\
C_0    &\simeq -1 + \sqrt{3-\lambda} + \frac{\epsilon}{2\sqrt{3-\lambda}}.\label{eqn:estimation4}
\end{align}
By choosing $\mu^2$, which sets the scale of the Polonyi field, one obtains $\nicefrac{V_0}{\mu^4}\sim\epsilon$, implying that the system under consideration can be used to construct a Minkowski minimum (or adjusting the vacuum energy to a small value). An example is presented in figs.\,\ref{fig:sdirection} and \ref{fig:3dplot}.

\subsection{Supersymmetry breaking parameters}
\label{sub:susybreaking}

Let us now have a look at the gravitino mass. It is given by
\begin{align}
m_{3/2} &= e^{G/2} = e^{K/2} \abs{W(S_0,C_0)} \simeq \mu^2,
\label{eqn:combinedgravitino}
\end{align}
and the Polonyi part in the superpotential dominates. So we see that $\mu^2$ controls not only the mass of the Polonyi field but also the gravitino mass. Furthermore, for the dilaton we obtain
\begin{align}
F^S &= e^{G/2}\, K^{-1}_{S\ol{S}}\, G_{\ol{S}}.
\label{eqn:combinedfs}
\end{align}
The quantity $G_{\ol{S}}$ can be estimated as follows. The stationary point conditions for the scalar potential eq.\,(\ref{eqn:newpot}) read
\begin{align}
V_S &= G_S\, V + e^{G} \frac{\del}{\del S}\klammer{K^{-1}_{S\ol{S}}G_S\, G_{\ol{S}}} + e^{G} \frac{\del}{\del S}\klammer{ G_C\, G_{\ol{C}} }\stackrel{!}{=}0, \label{eqn:stationarypoints}\\
V_C &= G_C\, V + e^{G}K^{-1}_{S\ol{S}}\, \frac{\del}{\del C}\klammer{G_S\, G_{\ol{S}}} + e^{G} \frac{\del}{\del C}\klammer{G_C\, G_{\ol{C}}}\stackrel{!}{=}0.\label{eqn:stationarypointc}
\end{align}
The first term in each equation vanishes at the minimum and the remaining terms give relations among the derivatives of $G$. The solution is given by
\begin{align}
G_{\ol{S}} \sim \frac{K_{S\ol{S}}\, G_C}{a},
\label{eqn:approximationgs}
\end{align}
with $G_C\sim\mathcal{O}(1)$. We then arrive at
\begin{align}
F^S \sim \frac{m_{3/2}}{a}.
\label{eqn:approximationfs}
\end{align}
The choice of a Minkowski vacuum (or small vacuum energy)
requires
$G_S$ to have a value 
that cancels the contribution of $K^{-1}_{S\ol{S}}$ 
in eq.\,(\ref{eqn:combinedfs}) 
and furthermore acts to suppress $F^S$ by $a$, where
\begin{align}
a\,\real{S} \sim \log\left(\frac{A}{\mu^2}\right) \sim \log\left(\frac{M_\s{Planck}}{m_{3/2}}\right)\sim \mathcal{O}(4\pi^2).
\label{eqn:suppression}
\end{align}
For the Polonyi field we have
\begin{align}
F^C &= e^{G/2}\, G_{C} \sim m_{3/2},
\label{eqn:approximationfc}
\end{align}
since $G_C\sim\mathcal{O}(1)$. Thus $F^C$ turns out to be the dominant part in SUSY breakdown whereas $F^S$ appears as being suppressed
by a factor as given in eq.\,(\ref{eqn:suppression}). 
The mass of the dilaton field is expressed as
\begin{align}
m^2_{S} &= \frac{V_{S\ol{S}}}{K_{S\ol{S}}},
\label{eqn:dilatonmass}
\end{align}
where the second derivative of the scalar potential has the behavior $V_{S\ol{S}}\sim a^2\, m^2_{3/2}$ and thus
\begin{align}
m^2_{S} \sim a^2\,m^2_{3/2}
\label{eqn:approximationms}
\end{align}
and the mass of the dilaton is enhanced compared to $m_{3/2}$. The mass of the Polonyi field is
\begin{align}
m^2_{C} &= \frac{V_{C\ol{C}}}{K_{C\ol{C}}} = V_{C\ol{C}} \sim m^2_{3/2},
\label{eqn:polonyimass}
\end{align}
which means that it is comparable to the gravitino mass. This is in 
accordance with the previous discussion, namely $\mu^2$ sets the scale of the Polonyi field as well as that of the gravitino mass.

\subsection[The little hierarchy: $\log{(M_\s{Planck}/{m_{3/2}})}$]{The little hierarchy: $\boldsymbol{\log{(M_\s{\bf Planck}/{m_{3/2}})}}$}
\label{sub:littlehierarchy}

If we compare this to the type IIB case we, in fact, end up
with very similar results (although the starting point
was quite different). In type IIB one started 
with a supersymmetric theory in an anti-de Sitter vacuum which then gets 
up-lifted to a Minkowski vacuum with SUSY breakdown induced by the 
up-lifting sector. In the heterotic case we started with 
an unstabilized dilaton. The superpotential interaction involving 
matter fields provides the stabilization of the dilaton and also  
induces the breakdown of supersymmetry. Finally a Minkowski
vacuum can be realized by properly adjusting the parameters 
of the Polonyi sector. The hierarchical structure among $m_S$, 
$m_{3/2}$ and gaugino masses $m_{1/2}\sim\nicefrac{F^S}{S+\ol{S}}$ is
\begin{align}
m_{S} \sim a\cdot m_{3/2} \sim a^2\cdot m_{1/2}.
\label{eqn:hierarchy}
\end{align}
This is similar as the result obtained in the type IIB case 
\cite{Choi:2004sx,Choi:2005ge}. In both cases, this little
hierarchy has its origin in the appearance of the factor
$\log{(M_\s{Planck}/{m_{3/2}})}$. It suppresses the modulus
contribution to the soft mass terms such that loop induced effects
become competitive to the tree level ones. The result is
the so-called mirage pattern of soft mass terms.

\section{Phenomenological properties of the set-up}
\label{sec:pheno}

Having presented the set-up we would like to analyze in detail the pattern 
of the emerging soft terms in the effective low energy theory. 
First we derive the soft breaking terms as boundary conditions 
valid at the grand unified theory (GUT) scale with a suitable parameterization. 
We then evolve these soft terms down to the electroweak (EW) 
scale and impose several phenomenological constraints of 
theoretical and experimental nature.

\subsection{Soft supersymmetry breaking terms}
\label{sub:softterms}

Our analysis in section~\ref{sub:combination} was done in the framework 
of the Polonyi model. Following the lines of \cite{Lebedev:2006qq} one 
can also construct generalized superpotentials for which the soft terms 
take a simpler form. In particular we are interested in a
simple pattern for the
gaugino masses. For a more general discussion see \cite{Choi:2007ka}.
In order to simplify the discussion we thus assume that
the minimum of the scalar potential emerges at $S_0=2$ and $C_0\ll 1$.
The full K\"ahler potential including interaction with observable 
matter fields is given by
\begin{align}
K &= -\log\klammer{S+\ol{S}} + C\ol{C} + Q_i\ol{Q}_i \left[\, 1 + \xi_i C\ol{C} \,\right],
\label{eqn:fullkahler}
\end{align}
where $Q_i$ are the visible sector matter fields and $\xi_i$ describe the coupling between observable and hidden matter. Furthermore we assume that the string threshold corrections to the gauge kinetic function involving the $C$ field are negligible at $C\ll 1$.  The moduli/dilaton mediated contribution to the soft terms is then given by
\begin{align}
M_a &= \frac{1}{2\real{f_a}}F^\alpha\del_\alpha f_a , \label{eqn:rawbc1} \\
A_{ijk} &= F^\alpha \Big[ K_\alpha + \del_\alpha\log Y_{ijk} -\del_\alpha\log\klammer{\mathcal{K}_i\mathcal{K}_j\mathcal{K}_k} \Big] , \label{eqn:rawbc2} \\
m^2_i &= m^2_{3/2} - \ol{F}^{\ol{\alpha}} F^\beta \del_{\ol{\alpha}}\del_\beta \log\mathcal{K}_i , \label{eqn:rawbc3}
\end{align}
where $\alpha$ and $\beta$ run over the SUSY breaking fields, $f_a=S$ are the gauge kinetic functions, $K_\alpha=\del_\alpha K$ and $\mathcal{K}_i$ is the K\"ahler metric for the visible fields
\begin{align}
\mathcal{K}_i &= \frac{\del^2 K}{\del Q_i \del \ol{Q}_i}.
\label{eqn:visiblemetric}
\end{align}
Assuming that $Y_{ijk}$ are independent of $S$ and $C$ we obtain
\begin{align}
M_a &= \frac{F^S}{S_0+\ol{S}_0}, \label{eqn:halfbc1} \\
A_{ijk} &= -\frac{F^S}{S_0+\ol{S}_0}, \label{eqn:halfbc2} \\
m^2_i &= m^2_{3/2} - \xi_i \abs{F^C}^2, \label{eqn:halfbc3}
\end{align}
with $K_S$ being the derivative of $K$ with respect to $S$. The condition for
 having a Minkowski vacuum gives a relation among $F^S$ and $F^C$, namely,
\begin{align}
\frac{\abs{F^S}^2}{\klammer{S_0+\ol{S}_0}^2} + \abs{F^C}^2 &= 3 m^2_{3/2}.
\label{eqn:ftermsconstraint}
\end{align}
Under these assumptions  (minimum at $C_0\ll 1$, $f_a$ 
independent of $C$),
the tree level soft terms for the gauginos and 
the $A$ parameters are independent 
of $F^C$. As discussed in section~\ref{sub:susybreaking}, 
$F^S$ is suppressed. 
Thus, we have to worry about loop suppressed 
contributions to the soft terms coming e.\,g. from the superconformal 
anomaly \cite{Randall:1998uk}. 
The masses squared of the scalars, however, could behave differently. 
They do contain contributions from $F^C$ which dominate over $F^S$. 
Nevertheless the 
$F^S$ contribution as well as the anomaly part may be of interest 
if we consider $\xi_i\sim\mathcal{O}(\nicefrac{1}{3})$. Including 
anomaly mediated contributions into eqs.\,(\ref{eqn:rawbc1}-\ref{eqn:rawbc3}) 
the GUT scale boundary values are given by
\begin{align}
M_a &= \frac{F^S}{S_0+\ol{S}_0} + b_a\, g^2_a \frac{F^\phi}{16\pi^2} , \label{eqn:metabc1} \\
A_{ijk} &= -\frac{F^S}{S_0+\ol{S}_0} + \klammer{\gamma_i+\gamma_j+\gamma_k}\frac{F^\phi}{16\pi^2} ,\label{eqn:metabc2} \\
m^2_i &= \xi_i \frac{\abs{F^S}^2}{(S_0+\ol{S}_0)^2} -\dot{\gamma}_i\frac{\abs{F^\phi}^2}{(16\pi^2)^2} +\frac{2 F^\phi F^S}{16\pi^2}\del_S \gamma_i + \klammer{1-3\xi_i}m^2_{3/2}, \label{eqn:metabc3}
\end{align}
with $F^\phi$ being the auxiliary field of the conformal compensator, $b_a$ are the beta function coefficients, $\gamma_i$ gives the anomalous dimension and $\dot{\gamma}_i=16\pi^2(\del\gamma_i/\del\log Q)$ with $Q$ being the renormalization scale. More details can be found in \cite{Choi:2005ge}.

For the gauginos eq.\,(\ref{eqn:metabc1}) we obtain a similar result as in the type IIB picture. They are split at the GUT scale according to their beta function coefficients. Since the evolution of the gauginos from the GUT to the EW scale is governed by the same beta function coefficients the splitting disappears at an intermediate scale, leading to the \emph{mirage} unification of the 
gaugino masses.

The form of the $A$ terms and the scalar masses is very similar 
to the type IIB case. For $\xi_i\sim0$, $F^C$ dominates and we obtain 
matter dominated SUSY breaking with $m^2_i\sim m^2_{3/2}$.
The choice $\xi_i\sim\mathcal{O}(\nicefrac{1}{3})$ would suppress the 
$F^C$ contribution and make it comparable to the dilaton and anomaly 
mediated contributions. The problematic feature of anomaly mediation is the 
potential presence of tachyonic sleptons. Due to the mixing between the 
dilaton and anomaly mediated contributions also the squarks 
might become tachyonic here. 
The presence of tachyons can be avoided  by appropriately choosing $\xi_i$.

As already studied in the type IIB case, the phenomenology of 
such a mixed mediation scheme depends crucially on the ratio between 
modulus/dilaton and anomaly contributions. 
We introduce the parameterization
\begin{align}
\varrho &\equiv \frac{1}{M}\,\frac{F^S}{S_0+\ol{S}_0}, \label{eqn:parameterrho} \\
M &\equiv \frac{m_{3/2}}{16\pi^2}, \label{eqn:parameterm}
\end{align}
where $\varrho$ measures the relative importance of dilaton and 
anomaly mediation and $M$ sets the scale of the soft mass 
terms. Note that the limit $\varrho=0$ corresponds 
to pure anomaly mediation whereas $\varrho\gg 1$ is pure dilaton 
domination. The last term in eq.\,(\ref{eqn:metabc3}) is enhanced 
by $(16\pi^2)^2$ with respect to the other terms. In order to compare its 
contribution with the remaining terms we use
\begin{align}
\eta^2_i &\equiv (1-3\xi_i)(16\pi^2)^2, \label{eqn:parametereta}
\end{align}
with $\eta_i=0$ corresponding to $\xi_i=\nicefrac{1}{3}$. Then, the soft terms eqs.\,(\ref{eqn:metabc1}-\ref{eqn:metabc3}) take the form
\begin{align}
M_a     &= M^{\phantom{2}} \Big[ \varrho + b_a g^2_a \Big], \label{eqn:bc1} \\
A_{ijk} &= M^{\phantom{2}} \Big[ -\varrho + \klammer{\gamma_i + \gamma_j + \gamma_k} \Big], \label{eqn:bc2} \\
m^2_i   &= M^2 \Big[ \xi_i\varrho^2 - \dot{\gamma_i} + 2\varrho\klammer{S_0+\ol{S}_0}\del_S\gamma_i + \eta^2_i \Big]. \label{eqn:bc3}
\end{align}
Let us close this section by pointing out that in the heterotic case 
the anomaly mediated contributions to the $A$ parameters and to the scalar 
masses squared (for $\eta_i\neq0$) are enhanced compared to the 
type IIB situation, where the modulus mediated contribution contained a 
factor of 3 originating from $3\log(T+\ol{T})$ as compared to $\log(S+\ol{S})$.

\subsection{Analysis of the parameters}
\label{sub:analysis}

The soft terms in our set-up are described by two continuous parameters
\begin{align}
\left\lbrace \varrho,\, m_{3/2}\right\rbrace ,
\end{align}
the three quantities
\begin{align}
\left\lbrace \tan\beta,\, \fkt{sign}\mu,\, m_t \vphantom{\varrho+m_{3/2}}\right\rbrace,
\end{align}
and the $\eta_i$ parameters from the matter sector. In the following
 we will assume non-universal masses for sfermions and Higgses 
\cite{Matalliotakis:1994ft,Olechowski:1994gm}
and denote
\begin{align}
\eta^{(\texttt{sfermions})}_i &\equiv \eta ,\\
\eta^{(\texttt{Higgses})}_i &\equiv \eta^\prime.
\end{align}
We use $\tan\beta$ to fix the $B\mu$ term. The requirement of correct electroweak symmetry breaking fixes the size of $\mu^2$ so its sign remains a free parameter. We will use $m_t=\unit[175]{GeV}$ for the top quark mass and $\fkt{sign}\mu=+1$ throughout our low energy analysis. For the calculation of the low energy data we use the public codes \texttt{SOFTSUSY} \cite{Allanach:2001kg} and \texttt{micrOMEGAs} \cite{Belanger:2001fz}.

\subsubsection{Gauginos}
\label{sub:gauginos}

The gaugino soft terms eq.\,(\ref{eqn:bc1}) approximately 
read 
\begin{align}
M_1 \simeq (3.3 + \varrho) M, \quad M_2 \simeq (0.5 + \varrho) M, \quad M_3 \simeq (-1.5 + \varrho) M.
\label{eqn:gutgauginos}
\end{align}
The non-universality of the gaugino masses arises from the anomaly 
mediated contributions which are proportional to the beta function 
coefficients $b_a = (\nicefrac{33}{5},1,-3)$. At the GUT scale the 
gauginos show a pattern  $M_1>M_2>M_3$ because $M_3$ 
is suppressed by the  
negative contribution from anomaly mediation.
Depending on the value of
$\varrho$  this negative contribution 
to $M_3$ might become more or less important.
At $\varrho\sim 1.5$ it might even lead to vanishing gluino mass.

\subsubsection{Scalars}
\label{sub:scalars}

\begin{figure}[t]
\begin{center}
\includegraphics[width=0.9\linewidth]{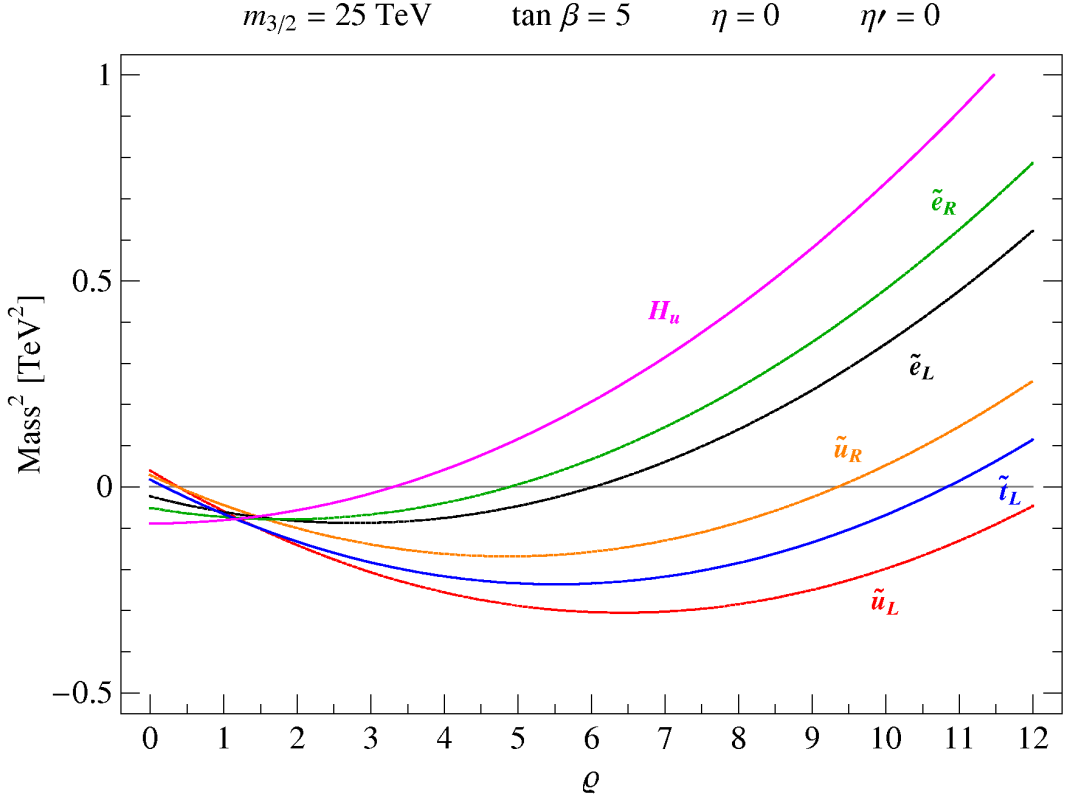}
\caption{Scalar squared masses at the GUT scale as functions of the ratio $\varrho$. For $\eta=\eta^\prime=0$ the  squarks as well as sleptons in the particular range of interest\newline $0\leq\varrho\leq 12$ appear to be tachyonic implying that the GUT scale boundary conditions eqs.\,(\ref{eqn:bc1}-\ref{eqn:bc3}) 
might be ill-defined. Introducing nonzero $\eta$ and $\eta^\prime$ provides positive contribution to the scalar masses squared.}\label{fig:inputscalars}
\end{center}
\end{figure}
As already analyzed in the type IIB case the scalar masses squared  
could become tachyonic due to the contributions from anomaly mediation. 
In pure anomaly mediation sleptons are tachyonic. Due to 
the mixing between dilaton and anomaly mediation also squarks might
become tachyonic here (fig.\,\ref{fig:inputscalars}). For small 
$\varrho$ the dilaton mediated contribution is too weak  
to cancel the negative anomaly contribution. Nevertheless 
we can use the $\eta$ and $\eta^\prime$ parameters to avoid this problem.
In oder to study a tachyon-free set-up for all values of $\varrho$ 
we scan over $\left\lbrace\eta,\eta^\prime\right\rbrace$ and exclude 
tachyonic regions. This implies lower bounds on $\eta\geq 3.5$ and
$\eta^\prime\geq 1.7$.

In fact, me might choose
the values of $\eta$ and $\eta^\prime$ in such a way that
the so-called ``MSSM hierarchy problem'' can be avoided. 
Correct electroweak symmetry breakdown (EWSB) requires
\begin{align}
\frac{m^2_Z}{2} &= -\mu^2 + \frac{m^2_{H_d} - m^2_{H_u}\tan^2\beta}{\tan^2\beta -1}
\label{eqn:ewsbcondition}
\end{align}
at the EW scale. Using the input soft terms eqs.\,(\ref{eqn:bc1}-\ref{eqn:bc3}) this condition can be rewritten at one-loop level as \cite{Kane:2002ap}
\begin{align}
\nonumber m^2_Z &\simeq -1.8\mu^2 + 5.9 M^2_3 - 0.4M^2_2 - 1.2 m^2_{H_u} + 0.9m^2_{Q_3} + 0.7m^2_{U_3} - 0.6A_t M_3 + 0.4 M_2 M_3\\
 &:= -1.8\mu^2 + \widetilde{m}^2_Z,
\label{eqn:oneloopcondition}
\end{align}
where we have considered $\tan\beta=5$ and neglected terms with 
smaller numerical coefficients. If all the parameters on the 
right hand side of eq.\,(\ref{eqn:oneloopcondition}) are of order 
of magnitude 
of \unit[100]{GeV}, no significant fine-tuning is needed. However, 
the soft masses are typically in the TeV region. Then, in order 
to obtain the correct value of $m_Z$,  cancellations in 
eq.\,(\ref{eqn:oneloopcondition}) are required. One could, of 
course, adjust $\mu$ such that $m_Z$ has the correct value but 
then $\mu$ might have to be very large. But one might also be
interested in a situation where $\mu$ has a value 
of order of the weak scale. This would then require cancellations 
inside $\widetilde{m}^2_Z$.

The largest contribution to $\widetilde{m}^2_Z$ comes from the 
gluino. In order to keep $\widetilde{m}^2_Z$ small one would have 
to keep $M_3$ under control
\cite{Choi:2005hd,Lebedev:2005ge}. 
As we have seen in section~\ref{sub:gauginos} the gluino mass 
might become small for small $\varrho$
(at $\varrho\sim1.5$ it might even vanish). This is ruled out, 
as the gluino would be the lightest supersymmetric particle (LSP).
In addition if the gluino 
is light it cannot provide the necessary renormalization group (RG) 
contribution to $m^2_{H_u}$ such that eq.\,(\ref{eqn:ewsbcondition}) 
will no longer be satisfied and consequently EWSB will not be 
realized around $\varrho\sim1.5$. 
Thus larger values of $\varrho$ are required. 
In order to achieve a cancellation within 
$\widetilde{m}^2_Z$ for moderate values of $\varrho$ one has 
to adjust the masses of the scalars (sfermions and Higgses). 
Here  the freedom of choosing $\eta$ and $\eta^\prime$ enters the
game.

As is evident from eq.\,(\ref{eqn:oneloopcondition}) the contribution 
from $m^2_{H_u}$ is negative and thus by increasing $m^2_{H_u}$ one 
obtains a sizeable term that could cancel  the contribution of the
gluino $M_3$. 
The contribution from squarks 
is positive and one has to keep their masses low. However, we cannot 
choose $\eta$ too small, otherwise the squarks might become tachyonic at 
the GUT scale. The essential lesson  we learn from these considerations 
is to keep $\eta$ as low as possible and then adjust $\eta^\prime$.
If we want to keep  $\widetilde{m}^2_Z\equiv(\unit[100]{GeV})^2$ a  
relation between the values of
$\varrho$, $m_{3/2}$, $\eta$ and $\eta^\prime$ 
has to be fulfilled. In that sense the ``MSSM hierarchy problem'' can
be avoided at the expense of a fine tuning of $\eta^\prime$.

\subsection{Constraints}
\label{sub:constraints}

After having excluded tachyons, we shall see that 
the parameter space of our set-up 
is further restricted by  
phenomenological constraints. These include the quest for correct
EWSB, mass bounds from LEP and
the cosmological relic abundance of neutralino dark matter.  
\begin{itemize}
\item \textit{Correct EWSB}\\
The minimization of the MSSM Higgs scalar potential leads 
to eq.\,(\ref{eqn:ewsbcondition}). Here $\mu^2$  
should be positive and $m^2_{H_u}$ should be negative at the 
EW scale. Given a positive $m^2_{H_u}$ at the GUT scale, it 
will be driven to negative values at the EW scale
by the renormalization group evolution according to
\begin{align}
\frac{d m^2_{H_u}}{d\log{Q}} \simeq \abs{y_t}^2 \left( m^2_{H_u} + m^2_{Q_3} + m^2_{U_3} \right) + \abs{A_t}^2.
\label{eqn:higgsrg}
\end{align}
The RG evolution is most sensitive to the gluino mass which induces
an increase of $m_{Q_3}$ and $m_{U_3}$. In a mirage mediation 
scheme as the one considered here
a cancellation between dilaton and anomaly mediated contributions for the 
gluino mass occurs for small values of $\varrho$
leading to an ultra-light 
gluino around $\varrho\sim1.5$. 
There the RG contribution from the 
gluino to eq.\,(\ref{eqn:higgsrg}) is too small and a satisfactory value of 
$m^2_{H_u}$ can not be obtained. 
{\emph{The requirement of correct EWSB
sets a lower bound on the $\varrho$ parameter.}}
\item \textit{LEP mass bounds}\\
Direct collider searches set lower bounds on the sparticle spectrum and Higgs masses. Most important and restrictive bounds are due to the lightest Higgs boson mass $m_h>\unit[114]{GeV}$, the lightest chargino $m_{\widetilde{\chi}^+}>\unit[103.5]{GeV}$ and the lightest stop quark $m_{\widetilde{t}_1}>\unit[95.7]{GeV}$ \cite{Yao:2006px}. Regions of parameter space violating one of these bounds are called \emph{below LEP}. 
{\emph{These constraints set a lower bound on $m_{3/2}$.}}
\item \textit{Neutralino Dark Matter}\\
In SUSY models the weakly interacting neutralinos  
tend to be the LSPs and 
they are perfect Dark Matter candidates
(under the assumption of $R$-parity conservation). 
In our model this 
is true throughout most of the parameter space. The four 
neutralinos of the MSSM $\widetilde{\chi}^0_{1,2,3,4}$ are 
superpositions of the neutral Higgs fermions  $\widetilde{H}^0_u$,  
$\widetilde{H}^0_d$ and the fermionic partners of the EW gauge 
bosons $\widetilde{B}^0$, $\widetilde{W}^0_3$. The neutralino 
mass matrix can be diagonalized by an orthogonal matrix $\mathcal{Z}$, 
such that the lightest neutralino is given by
\begin{align}
\widetilde{\chi}^0_1 = \mathcal{Z}_{11}\widetilde{B}^0 + \mathcal{Z}_{12}\widetilde{W}^0_3 + \mathcal{Z}_{13}\widetilde{H}^0_d + \mathcal{Z}_{14}\widetilde{H}^0_u.
\label{eqn:neutralinodecomp}
\end{align}
Using this decomposition one defines
\begin{align}
\widetilde{\chi}^0_1 = \left\{\begin{array}{cl}
                   \textrm{bino-like}     & \abs{\mathcal{Z}_{11}}^2+\abs{\mathcal{Z}_{12}}^2>0.9 \wedge \abs{\mathcal{Z}_{11}} > \abs{\mathcal{Z}_{12}}, \\
               \textrm{wino-like}     & \abs{\mathcal{Z}_{11}}^2+\abs{\mathcal{Z}_{12}}^2>0.9 \wedge \abs{\mathcal{Z}_{11}} < \abs{\mathcal{Z}_{12}} , \\
               \textrm{higgsino-like} & \abs{\mathcal{Z}_{11}}^2+\abs{\mathcal{Z}_{12}}^2<0.1, \\
           \textrm{mixed}         & \textrm{otherwise}.
                  \end{array}
            \right.
\label{eqn:lspnature}
\end{align}
We use the $3\sigma$ limit from the WMAP collaboration on the neutralino Dark Matter relic abundance \cite{Spergel:2006hy}:
\begin{align}
0.087 \leq \Omega_{\widetilde{\chi}}h^2 \leq 0.138.
\label{eqn:wmap}
\end{align}
We will require that the neutralinos annihilate efficiently enough 
to satisfy the bound eq. (\ref{eqn:wmap}) and assume that the LSP abundance 
is thermal. In the remainder of this paper the
regions of the parameter space that violate the upper WMAP 
bound are called \emph{forbidden}, those within the bounds are called  
\emph{favored} and those below the lower bound are denoted as  
\emph{allowed}. In the latter case the correct 
cosmological abundance of
Dark Matter could be achieved with additional dark matter particles
and/or a nonthermal origin.
\end{itemize}

\subsection{Low energy aspects of the spectrum}
\label{sub:mirage}

\begin{figure}[t]
\begin{center}
\includegraphics[width=0.9\linewidth]{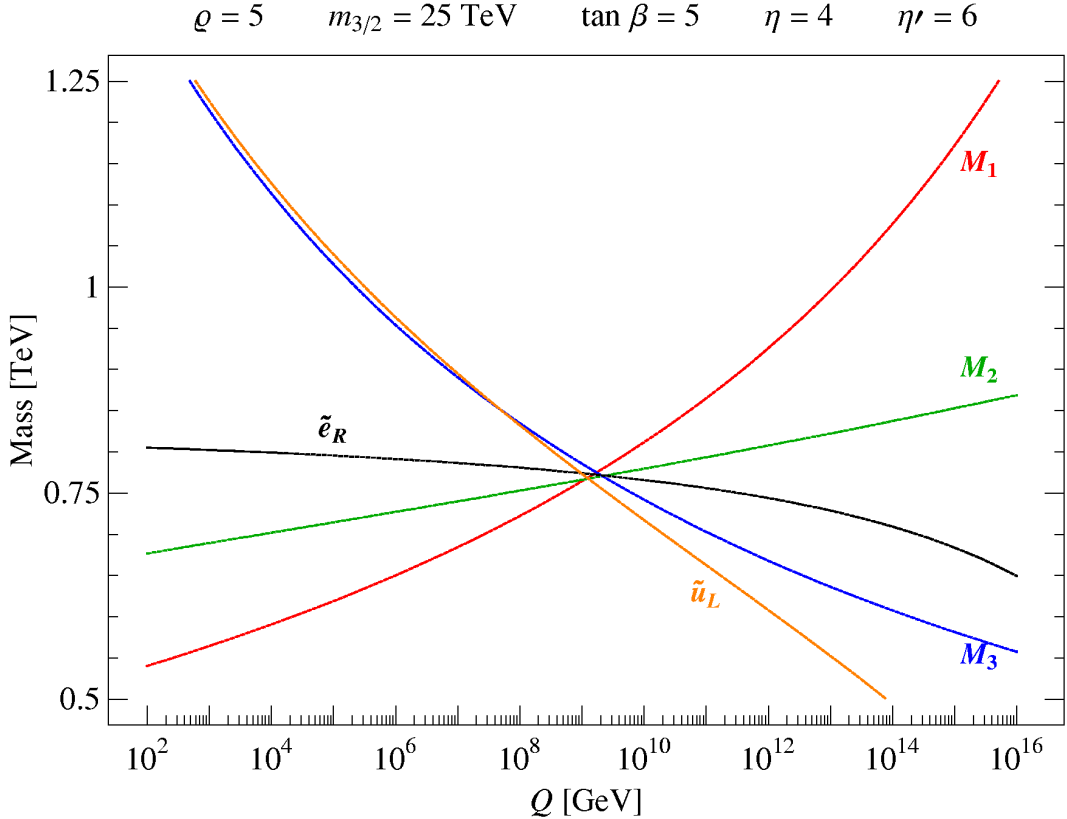}
\caption{Mirage unification of the gauginos and the first/second generation scalars.}\label{fig:mirage1}
\end{center}
\end{figure}
In models where the contributions from modulus and 
anomaly mediation are comparable we experience the phenomenon of 
mirage unification at an intermediate scale. 
The same happens, of course, in the present case of mixed 
dilaton-anomaly mediation.

\begin{itemize}
\item \textsl{Gauginos}\\
Below the GUT scale the nonuniversality of the gaugino masses is 
given by the respective beta function coefficients. The 
renormalization group running of the gaugino masses is governed by 
the same coefficients, thus at an intermediate scale, the 
splitting disappears and the gauginos unify. Since there 
is no physical threshold associated to this scale it is 
called \emph{mirage scale}. Using the parameterization 
eq.\,(\ref{eqn:parameterrho}) the mirage scale is given by
\begin{align}
M_\s{MIR} &= M_\s{GUT}\, e^{-\nicefrac{8\pi^2}{\varrho}}.
\label{eqn:miragescale}
\end{align}
For $\varrho=5$ the mirage scale is intermediate while 
for $\varrho\simeq2$ mirage unification occurs at the 
TeV scale. The pattern of the gaugino masses at the GUT scale 
is $M_1>M_2>M_3$. At the EW scale this pattern becomes 
inverted $M_3>M_2>M_1$ and a compressed spectrum is obtained (fig.\,\ref{fig:mirage1}).
\item \textsl{First and second generation scalars}\\
\begin{figure}[t]
\begin{center}
\includegraphics[width=0.9\linewidth]{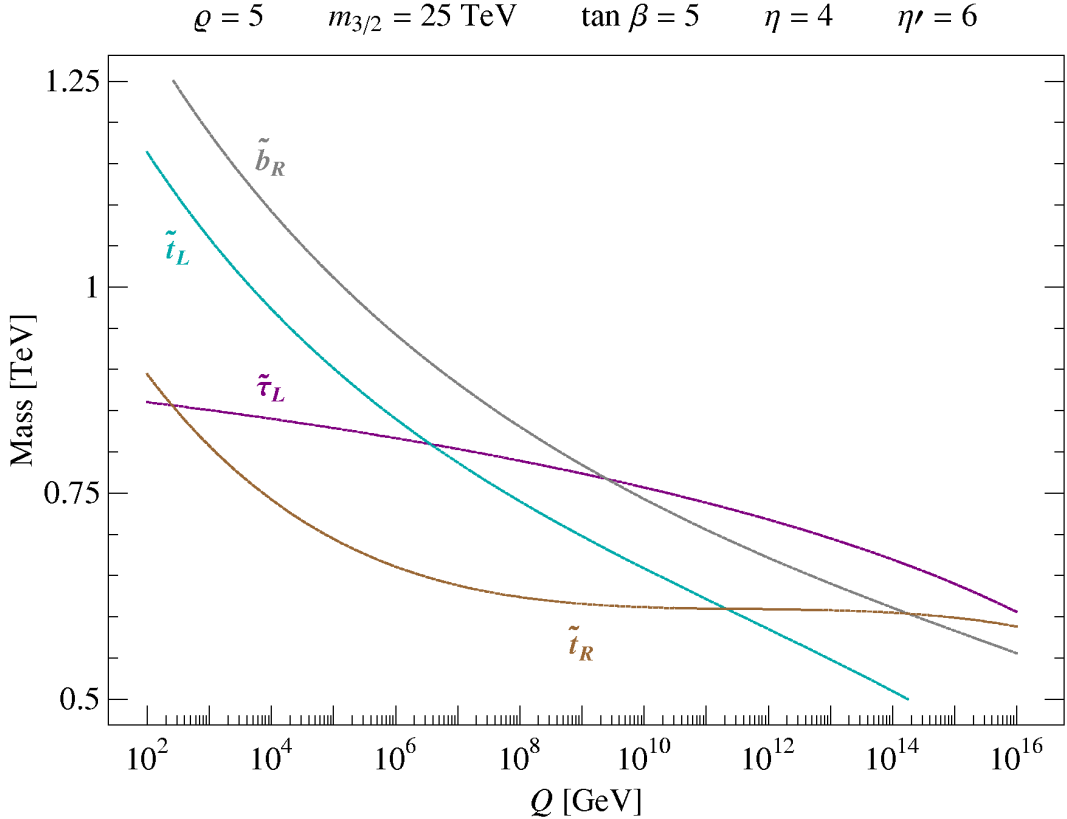}
\caption{RG evolution of the third generation scalars. Due to the 
influence of the Yukawa couplings these particle masses 
do not unify at the mirage scale. 
Nevertheless $\widetilde{b}_R$ and $\widetilde{\tau}_L$ seem to have 
a mirage unification point which is well approximated by 
eq.\,(\ref{eqn:miragescale}) and is robust (for low $\tan\beta$) 
under the variation of the parameters.}\label{fig:mirage2}
\end{center}
\end{figure}
These sparticles behave in a similar way as the gauginos. 
The reason for this is that they are unaffected by the 
large Yukawa coupling $y_t$ and string threshold 
corrections as well as K\"ahler anomalies are negligible. 
The renormalization group running of the first and second 
generation scalars is given by
\begin{align}
\frac{d m^2_i}{d\log{Q}} &\sim \sum_a g^2_a\, M^2_a\, \mathcal{C}^a_i.
\label{eqn:rgfirstsecond}
\end{align}
Under these circumstances the first and second generation scalars 
unify at the same $M_\s{MIR}$ eq.\,(\ref{eqn:miragescale}) as the 
gauginos (fig.\,\ref{fig:mirage1}).
\item \textsl{Third generation scalars}\\
Here we have to distinguish between those whose RG running depends 
on the Yukawa coupling $y_t$ 
and those that are unaffected by $y_t$.\footnote{This is strictly true only
for small values of $\tan\beta$. For large values of 
$\tan\beta$ we also have to take into account the bottom quark
Yukawa coupling $y_b$.} 
The only third generation scalars that feel the effect 
of $y_t$ are $m^2_{Q_3}$, $m^2_{U_3}$ and $m^2_{H_u}$. 
Consequently they do not unify at the mirage scale
eq.\,(\ref{eqn:miragescale}).
 All other third generation scalars and $m^2_{H_d}$ are only affected 
by the smaller Yukawa couplings $y_b, y_\tau \ll y_t$ and the 
structure of their RG running (fig.\,\ref{fig:mirage2}) is very similar to eq.\,(\ref{eqn:rgfirstsecond}), thus they (partially) share the mirage unification feature.
\end{itemize}

\subsubsection[Small $\tan\beta$ regime]{Small $\boldsymbol{\tan\beta}$ regime}
\label{sub:lowtb}

\begin{figure}[t]
\begin{center}
\includegraphics[width=0.9\linewidth]{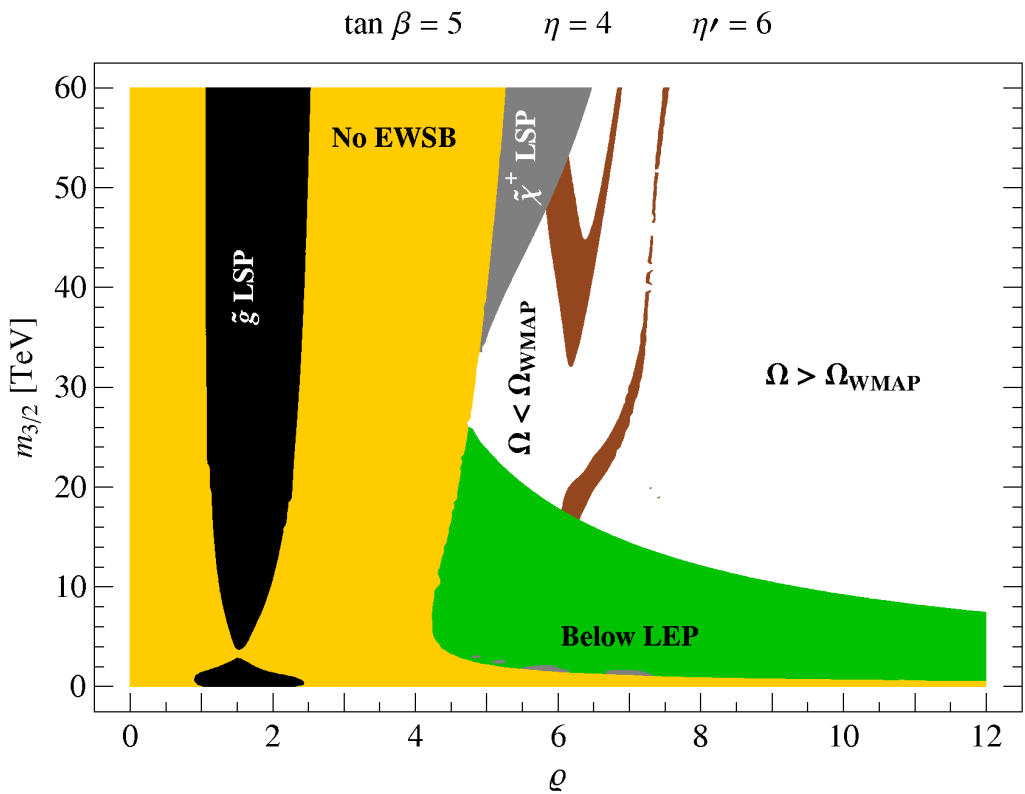}
\caption{Constraints on the parameter space $\{\varrho,m_{3/2}\}$ for $\tan\beta=5$ and\newline $\mu>0$. In the white region the low energy spectra are consistent with experimental and theoretical constraints. The brown strip satisfies the $3\sigma$ WMAP constraint. The constraint from 
$b\rightarrow s\gamma$ \cite{Chen:2001fja} does not further restrict
the allowed range of parameters.}\label{fig:lowtan}
\end{center}
\end{figure}
\begin{figure}[ht]
\begin{center}
\includegraphics[width=0.9\linewidth]{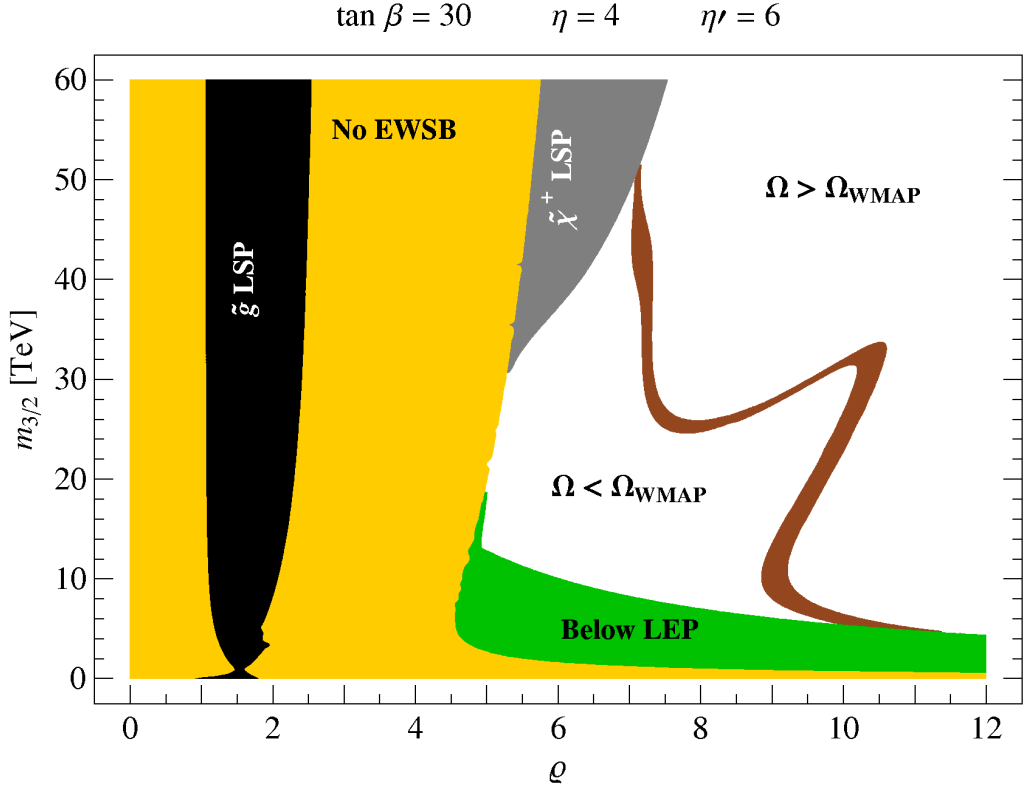}
\caption{Same as fig.\,\ref{fig:lowtan} but with $\tan\beta=30$. The LEP constraints are less restrictive whereas the chargino-LSP and the No-EWSB regions increase with larger $\tan\beta$. Compared to fig.\,\ref{fig:lowtan} larger portions of the parameter space are attractive for the discussion of the relic abundance.}\label{fig:hightan}
\end{center}
\end{figure}
As an illustrative example let us consider $\eta=4$ and $\eta^\prime=6$. The corresponding 
parameter space is shown in fig.~\ref{fig:lowtan}. In this scenario there are no tachyons. However, correct EWSB and current LEP bounds for $m_h$, $m_{\chargino}$ and $m_{\widetilde{t}}$ put severe constraints on $\varrho$ and $m_{3/2}$. Particularly we find that for $\tan\beta=5$ only $\varrho\geq 5$ and $m_{3/2}\geq\unit[8]{TeV}$ are allowed.

The presence of the no-EWSB region appears because at $\varrho\sim 1.5$ 
the gluino contribution to the RG is too small to make $m^2_{H_u}$ 
negative at the EW scale. In contrast to type IIB, in the heterotic case 
we find a region in the parameter space where a chargino is the LSP. 
This happens because the $A$ terms in the heterotic case are smaller 
than in type IIB.\footnote{More precisely, the  
contribution from modulus mediation in the $A$ terms is reduced by a 
factor of 3.} Additionally, these reduced $A$ terms lead to a smaller 
intra-generational mixing and increase the masses of the stop 
quark and the stau lepton. Therefore, stop or stau (N)LSP are
 not realized in this scenario.

For $\varrho$ values close to the no-EWSB region we have $\abs{\mu}<M_1$ and the neutralino LSP is higgsino-like. Going to larger $\varrho$ values the LSP becomes a mixed  higgsino-bino state. From $\varrho\sim7$ we have a mostly bino-like LSP and also $\abs{\mu}\gg M_1$. Thus, in most of the parameter space the LSP is bino-like.

The shaded/\cbox{brownochre} strip in fig.\,\ref{fig:lowtan} shows the region of the parameter space which is {\emph{favoured}} by the WMAP results eq.\,(\ref{eqn:wmap}). The region below the shaded/\cbox{brownochre} strip is \emph{allowed} (lower abundance) and that above (and to the right) of the shaded/\cbox{brownochre} strip is \emph{forbidden} (too large relic abundance). For $m_{3/2}=\unit[40]{TeV}$ and $\varrho\sim5$ we are close to the $\chargino$ LSP region and therefore we have $\widetilde{\chi}^0 \widetilde{\chi}^+$ coannihilation, which enhances the annihilation cross section and lowers the relic abundance. The neutralino in this region is higgsino-like. As $\varrho$ increases, the neutralino becomes mixed higgsino-bino and the $\mu$ term increases. The coannihilation with the chargino gets reduced and the annihilation cross section decreases leading to a higher relic abundance, so that there ($\varrho\sim6$) we reach the shaded/\cbox{brownochre} strip. When we proceed to increase $\varrho$, the 
$\widetilde{\chi}^0$ becomes bino-like. However, we then reach $m_A/2\sim m_{\widetilde{\chi}^0}$ and there the 
annihilation proceeds  efficiently through the pseudo-scalar Higgs exchange $\neutralino\neutralino\rightarrow A \rightarrow f\overline{f}$. This enhances the annihilation cross section and reduces the relic abundance. For $\varrho > 7$ the mass gap $\abs{m_A/2-m_{\neutralino}}$ grows and the efficiency of the $A$-channel reduces. As there are no other coannihilation channels available the cross section decreases and  the relic abundance becomes too large.

\subsubsection[Large $\tan\beta$ regime]{Large $\boldsymbol{\tan\beta}$ regime}
\label{sub:largetb}

For large values of $\tan\beta$, the LEP mass constraint becomes less restrictive (fig.~\ref{fig:hightan}). However, the no-EWSB region gets slightly bigger and the $\chargino$ LSP region covers a larger part of the parameter space (compared to the case of small $\tan\beta$). The composition of the neutralino LSP is similar to the $\tan\beta=5$ situation. For low $\varrho$ values (close to the no-EWSB region) the neutralino is higgsino-like. Then, for larger $\varrho$ it becomes more and more bino-like.

The shaded/\cbox{brownochre} strip, satisfying the WMAP limits, differs significantly from the one discussed above. 
Now, a larger part of the parameter space is consistent with the correct amount of Dark Matter. This is because for large $\tan\beta$ the pseudo-scalar Higgs exchange provides a sizable contribution to the annihilation cross section. For $m_{3/2}=\unit[30]{TeV}$ and $\varrho\sim5.5$ we have $m_{\chargino}\sim m_{\neutralino}\sim\mu$ and chargino coannihilation enhances the annihilation cross section and lowers the relic abundance. When $\varrho$ increases, $\mu$ gets larger (the whole sparticle spectrum becomes heavier) and the $\neutralino$ becomes bino-like. For $\varrho>7$ the mass gap $\abs{m_{\chargino}-m_{\neutralino}}$ grows, thus the $\chargino\neutralino$ coannihilation channel no longer provides a sizable effect. As a result, the annihilation cross section decreases and the relic abundance grows above the upper WMAP bound. At the same time the mass of the $\neutralino$ approaches the value $m_A/2$ and thus the pseudo-scalar Higgs exchange begins to contribute. The cross section $\sigma(\neutralino\neutralino\rightarrow A \rightarrow b\overline{b})$ grows with $\tan^2\beta$ and so this channel overcomes the decrease of the annihilation cross section caused by the bino component of the neutralino. For $\varrho>9.5$ the production of the relic abundance is in the allowed range. For still larger $\varrho$ finally this effect dies out and the relic abundance becomes too large.

\subsubsection{Numerical results}
Some points are selected from the the allowed parameter space 
in figs.\,\ref{fig:lowtan} and \ref{fig:hightan} and the spectrum is analysed
in detail. Examples of the spectra are displayed in 
table \ref{tab:samplepoints}. 

\begin{table}[t]
\small
\begin{center}
\begin{tabular}{w{2.5cm}z{1.2cm}z{1.2cm}z{1.2cm} }
\midrule\addlinespace
& \textbf{A} & \textbf{B}  & \textbf{C}  \\\addlinespace
$\tan\beta$             &      5  &     30  &    10   \\
$\varrho$               &      6  &     10  &    6.5  \\
$m_{3/2}$               &     40  &      6  &    25   \\
$\eta$                  &      4  &      4  &    5    \\
$\eta^\prime$           &      6  &      6  &    5    \\ \addlinespace\midrule
$M_1$                   &  1.040  &  0.211  &  0.676  \\
$M_2$                   &  1.317  &  0.311  &  0.881  \\
$M_3$                   &  2.391  &  0.743  &  1.697  \\
\cmidrule(r){1-1}\cmidrule(l){2-4}
$m_h$                   &  0.119  &  0.115  &  0.120  \\
$m_A$                   &  2.118  &  0.468  &  1.408  \\
$\mu$                   &  0.860  &  0.413  &  0.885  \\
\cmidrule(r){1-1}\cmidrule(l){2-4}
$m_{\tilde{\chi}^0_1}$  &  0.850  &  0.204  &  0.665  \\
$m_{\tilde{\chi}^0_2}$  &  0.870  &  0.296  &  0.838  \\
$m_{\tilde{\chi}^+_1}$  &  0.855  &  0.296  &  0.836  \\
\cmidrule(r){1-1}\cmidrule(l){2-4}
$m_{\tilde{t}_1}$       &  1.610  &  0.488  &  1.236  \\
$m_{\tilde{t}_2}$       &  2.110  &  0.694  &  1.578  \\
\cmidrule(r){1-1}\cmidrule(l){2-4}
$m_{\tilde{\tau}_1}$    &  1.398  &  0.233  &  1.021  \\
$m_{\tilde{\tau}_2}$    &  1.522  &  0.347  &  1.112  \\
\cmidrule(r){1-1}\cmidrule(l){2-4}
$\Omega_{\widetilde{\chi}}h^2$ & 0.088 & 0.115 & 0.092 \\
\midrule\midrule
\end{tabular}
\end{center}
\caption{Three sample spectra. All masses are given in TeV.}
\label{tab:samplepoints}
\end{table}

\section{Conclusions}

As we have seen, dilaton stabilization in the framework of
the heterotic string can be achieved quite easily, if we accept
the existence of an up-lifting sector (as postulated previously
in type IIB theory) which in any case is needed to adjust the
vacuum energy to an acceptable value. One just needs 
a gaugino condensate, while nontrivial background flux and/or
nonperturbative corrections to the K\"ahler potential are
not necessarily required. In that sense the heterotic 
mechanism of dilaton stabilization is somewhat similar to the 
one conjectured in the framework of M theory on $G_2$-manifolds
(which requires a racetrack scenario and an up-lifting sector
\cite{Acharya:2006ia,Acharya:2007rc}). A comparison of these
two scenarii will be the subject of future work, where we shall
also investigate heterotic M theory with a gaugino condensate
\cite{Horava:1996vs,Nilles:1997cm}.

Such a (partially sequestered) up-lifting sector is thus common
to many of the string schemes considered so far. While its
exact origin has to be clarified (branes, antibranes etc.),
its importance for the resulting phenomenology cannot be
overestimated. It is the dominant source of supersymmetry
breakdown, but as it is (partially) sequestered it leads to
mediation schemes where tree-level moduli contributions
compete with loop-effects from the up-lifting sector. The
gravitino and the moduli fields become rather heavy, but
the soft terms of the MSSM particles are suppressed by 
a factor of the order of $\log(M_\s{Planck}/m_{3/2})$
\cite{Choi:2005ge,LoaizaBrito:2005fa}. These soft 
masses often show a characteristic pattern, known as the
\emph{mirage pattern}. It leads to a rather compressed spectrum of 
masses and seems to be especially robust in the case of
gauginos \cite{Choi:2007ka}. This specific pattern has been investigated 
thoroughly in the framework of the type IIB string. Given the
particularly successful attempts of realistic MSSM model building
in the heterotic theory 
\cite{Lebedev:2006kn,Lebedev:2007hv,Kobayashi:2004ud,Forste:2004ie,Buchmuller:2005jr,Kim:2007mt}, 
it is reassuring to see that
a similar patterns seems to emerge here as well.

\vspace{0.6cm}
\subsection*{Acknowledgements}

We would like to thank Oleg Lebedev for valuable discussions. 
This work was partially supported by the
European Union 6th framework program MRTN-CT-2004-503069
``Quest for unification'', MRTN-CT-2004-005104 ``ForcesUniverse'',
MRTN-CT-2006-035863 ``UniverseNet'' and 
SFB-Transregio 33 ``The Dark Universe'' by Deutsche
Forschungsgemeinschaft (DFG). 
\vspace{0.6cm}

\addcontentsline{toc}{section}{{Bibliography}}

\begin{thebibliography}{10}

\bibitem{Lebedev:2006kn}
O.~Lebedev, H.~P. Nilles, S.~Raby, S.~Ramos-Sanchez, M.~Ratz, K.~S.
  Vaudrevange, and A.~Wingerter, ``{A mini-landscape of exact MSSM spectra in
  heterotic orbifolds},'' {\em Phys. Lett.} {\bf B645} (2007)  88--94,
\href{http://arxiv.org/abs/hep-th/0611095}{{\tt hep-th/0611095}}.

\bibitem{Lebedev:2007hv}
O.~Lebedev, H.~P. Nilles, S.~Raby, S.~Ramos-Sanchez, M.~Ratz, K.~S.
  Vaudrevange, and A.~Wingerter, ``{The Heterotic Road to the MSSM with R
  parity},'' \href{http://dx.doi.org/10.1103/PhysRevD.77.046013}{{\em Phys.
  Rev.} {\bf D77} (2008)  046013},
\href{http://arxiv.org/abs/0708.2691}{{\tt arXiv:0708.2691 [hep-th]}}.

\bibitem{Choi:2006qh}
K.-S. Choi and J.~E. Kim, ``{Quarks and leptons from orbifolded superstring},''
  {\em Springer: Lecture Notes in Physics} {\bf 696} (2006)  1.

\bibitem{Derendinger:1985kk}
J.~P. Derendinger, L.~E. Ibanez, and H.~P. Nilles, ``{On the Low-Energy d = 4,
  N=1 Supergravity Theory Extracted from the d=10, N=1 Superstring},''
{\em Phys. Lett.} {\bf B155} (1985)  65.

\bibitem{Dine:1985rz}
M.~Dine, R.~Rohm, N.~Seiberg, and E.~Witten, ``{Gluino Condensation in
  Superstring Models},''
{\em Phys. Lett.} {\bf B156} (1985)  55.

\bibitem{Derendinger:1985cv}
J.~P. Derendinger, L.~E. Ibanez, and H.~P. Nilles, ``{On the Low-Energy Limit
  of Superstring Theories},''
{\em Nucl. Phys.} {\bf B267} (1986)  365.

\bibitem{Krasnikov:1987jj}
N.~V. Krasnikov, ``{On Supersymmetry Breaking in Superstring Theories},''
{\em Phys. Lett.} {\bf B193} (1987)  37--40.

\bibitem{Casas:1990qi}
J.~A. Casas, Z.~Lalak, C.~Munoz, and G.~G. Ross, ``{HIERARCHICAL SUPERSYMMETRY
  BREAKING AND DYNAMICAL DETERMINATION OF COMPACTIFICATION PARAMETERS BY
  NONPERTURBATIVE EFFECTS},''
{\em Nucl. Phys.} {\bf B347} (1990)  243--269.

\bibitem{Casas:1996zi}
J.~A. Casas, ``{The generalized dilaton supersymmetry breaking scenario},''
  {\em Phys. Lett.} {\bf B384} (1996)  103--110,
\href{http://arxiv.org/abs/hep-th/9605180}{{\tt hep-th/9605180}}.

\bibitem{Binetruy:1996xja}
P.~Binetruy, M.~K. Gaillard, and Y.-Y. Wu, ``{Dilaton Stabilization in the
  Context of Dynamical Supersymmetry Breaking through Gaugino Condensation},''
  {\em Nucl. Phys.} {\bf B481} (1996)  109--128,
\href{http://arxiv.org/abs/hep-th/9605170}{{\tt hep-th/9605170}}.

\bibitem{Barreiro:1997rp}
T.~Barreiro, B.~de~Carlos, and E.~J. Copeland, ``{On non-perturbative
  corrections to the Kaehler potential},'' {\em Phys. Rev.} {\bf D57} (1998)
  7354--7360,
\href{http://arxiv.org/abs/hep-ph/9712443}{{\tt hep-ph/9712443}}.

\bibitem{Dasgupta:1999ss}
K.~Dasgupta, G.~Rajesh, and S.~Sethi, ``{M theory, orientifolds and G-flux},''
  {\em JHEP} {\bf 08} (1999)  023,
\href{http://arxiv.org/abs/hep-th/9908088}{{\tt hep-th/9908088}}.

\bibitem{Giddings:2001yu}
S.~B. Giddings, S.~Kachru, and J.~Polchinski, ``{Hierarchies from fluxes in
  string compactifications},'' {\em Phys. Rev.} {\bf D66} (2002)  106006,
\href{http://arxiv.org/abs/hep-th/0105097}{{\tt hep-th/0105097}}.

\bibitem{Becker:2003yv}
K.~Becker, M.~Becker, K.~Dasgupta, and P.~S. Green, ``{Compactifications of
  heterotic theory on non-Kaehler complex manifolds. I},'' {\em JHEP} {\bf 04}
  (2003)  007,
\href{http://arxiv.org/abs/hep-th/0301161}{{\tt hep-th/0301161}}.

\bibitem{LopesCardoso:2003sp}
G.~Lopes~Cardoso, G.~Curio, G.~Dall'Agata, and D.~Lust, ``{Heterotic string
  theory on non-Kaehler manifolds with H- flux and gaugino condensate},'' {\em
  Fortsch. Phys.} {\bf 52} (2004)  483--488,
\href{http://arxiv.org/abs/hep-th/0310021}{{\tt hep-th/0310021}}.

\bibitem{Gurrieri:2004dt}
S.~Gurrieri, A.~Lukas, and A.~Micu, ``{Heterotic on half-flat},'' {\em Phys.
  Rev.} {\bf D70} (2004)  126009,
\href{http://arxiv.org/abs/hep-th/0408121}{{\tt hep-th/0408121}}.

\bibitem{Kachru:2003aw}
S.~Kachru, R.~Kallosh, A.~Linde, and S.~P. Trivedi, ``{De Sitter vacua in
  string theory},'' {\em Phys. Rev.} {\bf D68} (2003)  046005,
\href{http://arxiv.org/abs/hep-th/0301240}{{\tt hep-th/0301240}}.

\bibitem{Choi:2004sx}
K.~Choi, A.~Falkowski, H.~P. Nilles, M.~Olechowski, and S.~Pokorski,
  ``{Stability of flux compactifications and the pattern of supersymmetry
  breaking},'' {\em JHEP} {\bf 11} (2004)  076,
\href{http://arxiv.org/abs/hep-th/0411066}{{\tt hep-th/0411066}}.

\bibitem{Choi:2005ge}
K.~Choi, A.~Falkowski, H.~P. Nilles, and M.~Olechowski, ``{Soft supersymmetry
  breaking in KKLT flux compactification},'' {\em Nucl. Phys.} {\bf B718}
  (2005)  113--133,
\href{http://arxiv.org/abs/hep-th/0503216}{{\tt hep-th/0503216}}.

\bibitem{Choi:2005uz}
K.~Choi, K.~S. Jeong, and K.-i. Okumura, ``{Phenomenology of mixed
  modulus-anomaly mediation in fluxed string compactifications and brane
  models},'' {\em JHEP} {\bf 09} (2005)  039,
\href{http://arxiv.org/abs/hep-ph/0504037}{{\tt hep-ph/0504037}}.

\bibitem{Endo:2005uy}
M.~Endo, M.~Yamaguchi, and K.~Yoshioka, ``{A bottom-up approach to moduli
  dynamics in heavy gravitino scenario: Superpotential, soft terms and
  sparticle mass spectrum},''
  \href{http://dx.doi.org/10.1103/PhysRevD.72.015004}{{\em Phys. Rev.} {\bf
  D72} (2005)  015004},
\href{http://arxiv.org/abs/hep-ph/0504036}{{\tt arXiv:hep-ph/0504036}}.

\bibitem{Falkowski:2005ck}
A.~Falkowski, O.~Lebedev, and Y.~Mambrini, ``{SUSY phenomenology of KKLT flux
  compactifications},'' {\em JHEP} {\bf 11} (2005)  034,
\href{http://arxiv.org/abs/hep-ph/0507110}{{\tt hep-ph/0507110}}.

\bibitem{Baer:2006id}
H.~Baer, E.-K. Park, X.~Tata, and T.~T. Wang, ``{Collider and dark matter
  searches in models with mixed modulus-anomaly mediated SUSY breaking},'' {\em
  JHEP} {\bf 08} (2006)  041,
\href{http://arxiv.org/abs/hep-ph/0604253}{{\tt hep-ph/0604253}}.

\bibitem{Lebedev:2006tr}
O.~Lebedev, H.~P. Nilles, S.~Raby, S.~Ramos-Sanchez, M.~Ratz, K.~S.
  Vaudrevange, and A.~Wingerter, ``{Low Energy Supersymmetry from the Heterotic
  Landscape},'' {\em Phys. Rev. Lett.} {\bf 98} (2007)  181602,
\href{http://arxiv.org/abs/hep-th/0611203}{{\tt hep-th/0611203}}.

\bibitem{Lebedev:2006qq}
O.~Lebedev, H.~P. Nilles, and M.~Ratz, ``{de Sitter vacua from matter
  superpotentials},'' {\em Phys. Lett.} {\bf B636} (2006)  126,
\href{http://arxiv.org/abs/hep-th/0603047}{{\tt hep-th/0603047}}.

\bibitem{GomezReino:2006dk}
M.~Gomez-Reino and C.~A. Scrucca, ``{Locally stable non-supersymmetric
  Minkowski vacua in supergravity},'' {\em JHEP} {\bf 05} (2006)  015,
\href{http://arxiv.org/abs/hep-th/0602246}{{\tt hep-th/0602246}}.

\bibitem{Lust:2006zg}
D.~Lust, S.~Reffert, E.~Scheidegger, W.~Schulgin, and S.~Stieberger, ``{Moduli
  stabilization in type IIB orientifolds. II},'' {\em Nucl. Phys.} {\bf B766}
  (2007)  178--231,
\href{http://arxiv.org/abs/hep-th/0609013}{{\tt hep-th/0609013}}.

\bibitem{Dudas:2006gr}
E.~Dudas, C.~Papineau, and S.~Pokorski, ``{Moduli stabilization and uplifting
  with dynamically generated F-terms},'' {\em JHEP} {\bf 02} (2007)  028,
\href{http://arxiv.org/abs/hep-th/0610297}{{\tt hep-th/0610297}}.

\bibitem{Abe:2006xp}
H.~Abe, T.~Higaki, T.~Kobayashi, and Y.~Omura, ``{Moduli stabilization, F-term
  uplifting and soft supersymmetry breaking terms},'' {\em Phys. Rev.} {\bf
  D75} (2007)  025019,
\href{http://arxiv.org/abs/hep-th/0611024}{{\tt hep-th/0611024}}.

\bibitem{Lebedev:2006qc}
O.~Lebedev, V.~Lowen, Y.~Mambrini, H.~P. Nilles, and M.~Ratz, ``{Metastable
  vacua in flux compactifications and their phenomenology},'' {\em JHEP} {\bf
  02} (2007)  063,
\href{http://arxiv.org/abs/hep-ph/0612035}{{\tt hep-ph/0612035}}.

\bibitem{Serone:2007sv}
M.~Serone and A.~Westphal, ``{Moduli Stabilization in Meta-Stable Heterotic
  Supergravity Vacua},'' {\em JHEP} {\bf 08} (2007)  080,
\href{http://arxiv.org/abs/arXiv:0707.0497 [hep-th]}{{\tt arXiv:0707.0497
  [hep-th]}}.

\bibitem{Intriligator:2006dd}
K.~Intriligator, N.~Seiberg, and D.~Shih, ``{Dynamical SUSY breaking in
  meta-stable vacua},'' {\em JHEP} {\bf 04} (2006)  021,
\href{http://arxiv.org/abs/hep-th/0602239}{{\tt hep-th/0602239}}.

\bibitem{Polchinski:1998rq}
J.~Polchinski, ``{String theory. Vol. 1: An introduction to the bosonic
  string},''. Cambridge, UK: Univ. Pr. (1998) 402 p.

\bibitem{Polonyi:1977pj}
J.~Polonyi, ``{Generalization of the Massive Scalar Multiplet Coupling to the
  Supergravity},''. Hungary Central Inst Res - KFKI-77-93 (77,REC.JUL 78) 5p.

\bibitem{Choi:2007ka}
K.~Choi and H.~P. Nilles, ``{The gaugino code},'' {\em JHEP} {\bf 04} (2007)
  006,
\href{http://arxiv.org/abs/hep-ph/0702146}{{\tt hep-ph/0702146}}.

\bibitem{Randall:1998uk}
L.~Randall and R.~Sundrum, ``{Out of this world supersymmetry breaking},'' {\em
  Nucl. Phys.} {\bf B557} (1999)  79--118,
\href{http://arxiv.org/abs/hep-th/9810155}{{\tt hep-th/9810155}}.

\bibitem{Matalliotakis:1994ft}
D.~Matalliotakis and H.~P. Nilles, ``{Implications of nonuniversality of soft
  terms in supersymmetric grand unified theories},'' {\em Nucl. Phys.} {\bf
  B435} (1995)  115--128,
\href{http://arxiv.org/abs/hep-ph/9407251}{{\tt hep-ph/9407251}}.

\bibitem{Olechowski:1994gm}
M.~Olechowski and S.~Pokorski, ``{Electroweak symmetry breaking with
  nonuniversal scalar soft terms and large tan beta solutions},'' {\em Phys.
  Lett.} {\bf B344} (1995)  201--210,
\href{http://arxiv.org/abs/hep-ph/9407404}{{\tt hep-ph/9407404}}.

\bibitem{Allanach:2001kg}
B.~C. Allanach, ``{SOFTSUSY: A C++ program for calculating supersymmetric
  spectra},'' {\em Comput. Phys. Commun.} {\bf 143} (2002)  305--331,
\href{http://arxiv.org/abs/hep-ph/0104145}{{\tt hep-ph/0104145}}.

\bibitem{Belanger:2001fz}
G.~Belanger, F.~Boudjema, A.~Pukhov, and A.~Semenov, ``{micrOMEGAs: A program
  for calculating the relic density in the MSSM},'' {\em Comput. Phys. Commun.}
  {\bf 149} (2002)  103--120,
\href{http://arxiv.org/abs/hep-ph/0112278}{{\tt hep-ph/0112278}}.

\bibitem{Kane:2002ap}
G.~L. Kane, J.~D. Lykken, B.~D. Nelson, and L.-T. Wang, ``{Re-examination of
  electroweak symmetry breaking in supersymmetry and implications for light
  superpartners},'' {\em Phys. Lett.} {\bf B551} (2003)  146--160,
\href{http://arxiv.org/abs/hep-ph/0207168}{{\tt hep-ph/0207168}}.

\bibitem{Choi:2005hd}
K.~Choi, K.~S. Jeong, T.~Kobayashi, and K.-i. Okumura, ``{Little SUSY hierarchy
  in mixed modulus-anomaly mediation},'' {\em Phys. Lett.} {\bf B633} (2006)
  355--361,
\href{http://arxiv.org/abs/hep-ph/0508029}{{\tt hep-ph/0508029}}.

\bibitem{Lebedev:2005ge}
O.~Lebedev, H.~P. Nilles, and M.~Ratz, ``{A note on fine-tuning in mirage
  mediation},''
\href{http://arxiv.org/abs/hep-ph/0511320}{{\tt hep-ph/0511320}}.

\bibitem{Yao:2006px}
{\bf {Particle Data Group}} Collaboration, W.~M. Yao {\em et al.}, ``Review of
  particle physics,''
{\em J. Phys.} {\bf G33} (2006)  1--1232.

\bibitem{Spergel:2006hy}
{\bf WMAP} Collaboration, D.~N. Spergel {\em et al.}, ``{Wilkinson Microwave
  Anisotropy Probe (WMAP) three year results: Implications for cosmology},''
  {\em Astrophys. J. Suppl.} {\bf 170} (2007)  377,
\href{http://arxiv.org/abs/astro-ph/0603449}{{\tt astro-ph/0603449}}.

\bibitem{Chen:2001fja}
{\bf CLEO} Collaboration, S.~Chen {\em et al.}, ``{Branching fraction and
  photon energy spectrum for $b\rightarrow s\gamma$},'' {\em Phys. Rev. Lett.}
  {\bf 87} (2001)  251807,
\href{http://arxiv.org/abs/hep-ex/0108032}{{\tt hep-ex/0108032}}.

\bibitem{Acharya:2006ia}
B.~Acharya, K.~Bobkov, G.~Kane, P.~Kumar, and D.~Vaman, ``{An M theory solution
  to the hierarchy problem},'' {\em Phys. Rev. Lett.} {\bf 97} (2006)  191601,
\href{http://arxiv.org/abs/hep-th/0606262}{{\tt hep-th/0606262}}.

\bibitem{Acharya:2007rc}
B.~S. Acharya, K.~Bobkov, G.~L. Kane, P.~Kumar, and J.~Shao, ``{Explaining the
  electroweak scale and stabilizing moduli in M theory},'' {\em Phys. Rev.}
  {\bf D76} (2007)  126010,
\href{http://arxiv.org/abs/hep-th/0701034}{{\tt hep-th/0701034}}.

\bibitem{Horava:1996vs}
P.~Horava, ``{Gluino condensation in strongly coupled heterotic string
  theory},'' {\em Phys. Rev.} {\bf D54} (1996)  7561--7569,
\href{http://arxiv.org/abs/hep-th/9608019}{{\tt hep-th/9608019}}.

\bibitem{Nilles:1997cm}
H.~P. Nilles, M.~Olechowski, and M.~Yamaguchi, ``{Supersymmetry breaking and
  soft terms in M-theory},'' {\em Phys. Lett.} {\bf B415} (1997)  24--30,
\href{http://arxiv.org/abs/hep-th/9707143}{{\tt hep-th/9707143}}.

\bibitem{LoaizaBrito:2005fa}
O.~Loaiza-Brito, J.~Martin, H.~P. Nilles, and M.~Ratz,
  ``$\log(M_{Pl}/m_{3/2}$),'' {\em AIP Conf. Proc.} {\bf 805} (2006)  198--204,
\href{http://arxiv.org/abs/hep-th/0509158}{{\tt hep-th/0509158}}.

\bibitem{Kobayashi:2004ud}
T.~Kobayashi, S.~Raby, and R.-J. Zhang, ``{Constructing 5d orbifold grand
  unified theories from heterotic strings},'' {\em Phys. Lett.} {\bf B593}
  (2004)  262--270,
\href{http://arxiv.org/abs/hep-ph/0403065}{{\tt hep-ph/0403065}}.

\bibitem{Forste:2004ie}
S.~Forste, H.~P. Nilles, P.~K.~S. Vaudrevange, and A.~Wingerter, ``{Heterotic
  brane world},'' {\em Phys. Rev.} {\bf D70} (2004)  106008,
\href{http://arxiv.org/abs/hep-th/0406208}{{\tt hep-th/0406208}}.

\bibitem{Buchmuller:2005jr}
W.~Buchmuller, K.~Hamaguchi, O.~Lebedev, and M.~Ratz, ``{Supersymmetric
  standard model from the heterotic string},'' {\em Phys. Rev. Lett.} {\bf 96}
  (2006)  121602,
\href{http://arxiv.org/abs/hep-ph/0511035}{{\tt hep-ph/0511035}}.

\bibitem{Kim:2007mt}
J.~E. Kim, J.-H. Kim, and B.~Kyae, ``{Superstring standard model from Z(12-I)
  orbifold compactification with and without exotics, and effective R-
  parity},'' {\em JHEP} {\bf 06} (2007)  034,
\href{http://arxiv.org/abs/hep-ph/0702278}{{\tt hep-ph/0702278}}.

\end{thebibliography}
\providecommand{\href}[2]{#2}\begingroup\raggedright\endgroup

\end{document}